\documentclass[final,5p,times]{elsarticle}
\usepackage{amsmath,amssymb,amsfonts}
\usepackage{siunitx}
\usepackage{wrapfig}
\usepackage{graphicx}
\usepackage{soul}
\usepackage{tikz}
\usepackage{multirow}
\usepackage[shortcuts,acronym]{glossaries}
\usepackage[english]{babel}

\newacronym{KPI}{KPI}{Key Performance Indicator}
\newacronym{NF}{NF}{Network Function}
\newacronym{VNF}{VNF}{Virtual Network Function}
\newacronym{MAPE-K}{MAPE-K}{Monitor-Analyze-Plan-Execute over a shared Knowledge}
\newacronym{QoE}{QoE}{Quality of Experience}
\newacronym{NI}{NI}{Network Intelligence}
\newacronym{RAT}{RAT}{Radio Access Technologies}
\newacronym{ML}{ML}{Machine Learning}
\newacronym{GRU}{GRU}{Gated Recurrent Units}
\newacronym{GB}{GB}{Gradient Boost}
\newacronym{DL}{DL}{Deep Learning}
\newacronym{DNN}{DNN}{Deep Neural Network}
\newacronym{NN}{NN}{Neural Network}
\newacronym{WLAN}{WLAN}{Wireless Local Area Network}
\newacronym{TC}{TC}{Traffic Classification}
\newacronym{QoS}{QoS}{Quality of Service}
\newacronym{OSI}{OSI}{Open Systems Interconnection}
\newacronym{NMS}{NMS}{Network Monitoring Service}
\newacronym{DPI}{DPI}{Deep Packet Inspection}
\newacronym{IP}{IP}{Internet Protocol}
\newacronym{VPN}{VPN}{Virtual Private Network}
\newacronym{AP}{AP}{Access Point}
\newacronym{CNN}{CNN}{Convolutional Neural Network}
\newacronym{RNN}{RNN}{Recurrent Neural Networks}
\newacronym{PHY}{PHY}{Physical-layer}
\newacronym{CR}{CR}{Cognitive Radio}
\newacronym{RAN}{RAN}{Radio Access Network}
\newacronym{vRAN}{vRAN}{Radio Access Network virtualization}
\newacronym{vRAP}{vRAP}{Virtual Radio Access Point}
\newacronym{RRU}{RRU}{Remote Radio Units}
\newacronym{BBU}{BBU}{Base Band Units}
\newacronym{SVM}{SVM}{Supported Vector Machine}
\newacronym{LTE}{LTE}{Long-Term Evolution}
\newacronym{TR}{TR}{Technology Recognition}
\newacronym{IRAN}{IRAN}{Intelligent RAN}
\newacronym{ImRAT}{ImRAT}{Intelligent multi-RAT}
\newacronym{UT}{UT}{User's Terminal}
\newacronym{AI}{AI}{Artificial Intelligence}
\newacronym{IQ}{IQ}{In-phase and Quadrature components}
\newacronym{FFT}{FFT}{Fast Fourier Transform }
\newacronym{STFT}{STFT}{Short-Time Fourier Transform }
\newacronym{CWT}{CWT}{Continuous Wavelet Transform}
\newacronym{WNIC}{WNIC}{Wireless Network Interface Card}
\newacronym{ICDE}{ICDE}{Intelligent Control and Decision Engine}
\newacronym{MCS}{MCS}{Modulation and Coding Scheme}
\newacronym{RKHS}{RKHS}{Reproducing Kernel Hilbert Space}
\newacronym{MLP}{MLP}{Multi-Layer Perceptron}
\newacronym{NB}{NB}{Naïve Bayes}
\newacronym{DT}{DT}{Decision Tree}
\newacronym{SDAE}{SDAE}{Stacked Denoising AutoEncoder}
\newacronym{RF}{RF}{Random Forest}
\newacronym{AE}{AE}{AutoEncoder}
\newacronym{LSTM}{LSTM}{Long Short-Term Memory}
\newacronym{TCP}{TCP}{Transmission Control Protocol}
\newacronym{GW}{GW}{Gateway}
\newacronym{CIR}{CIR}{Collaborative Intelligent Radio}
\newacronym{VAE}{VAE}{Variational AutoEncoder}
\newacronym{SAE}{SAE}{Stacked AutoEncoder}
\newacronym{DSSS}{DSSS}{Direct-Sequence Spread Spectrum}
\newacronym{OFDM}{OFDM}{Orthogonal Frequency Division Multiplexing}
\newacronym{GAN}{GAN}{Generative Adversarial Network}
\newacronym{k-NN}{k-NN}{k-Nearest Neighbors}
\newacronym{L1}{L1}{Physical Layer}
\newacronym{L2}{L2}{Link Layer}
\newacronym{L7}{L7}{Application Layer}
\newacronym{L3}{L3}{IP Layer}
\newacronym{Conv}{Conv}{Convolutional}
\newacronym{SNR}{SNR}{Signal-To-Noise-Ratio}
\newacronym{HDLC}{HDLC}{High-level Data Link Control}
\newacronym{QPSK}{QPSK}{Quadrature Phase Shift Keying}
\newacronym{QAM}{QAM}{Quadrature Amplitude Modulation}
\newacronym{BPSK}{BPSK}{Binary Phase Shift Keying}
\newacronym{CCK}{CCK}{Complementary Code Keying}
\newacronym{SDR}{SDR}{Software Defined Radios}
\newacronym{WPA}{WPA}{Wi-Fi Protected Access}
\newacronym{MTL}{MTL}{Multi-Task Learning}
\newacronym{DCI}{DCI}{Downlink Control Information}
\newacronym{LTE-PDCCH}{LTE-PDCCH}{LTE Physical Downlink Control CHannel}
\newacronym{GL}{GL}{Gossip Learning}
\newacronym{ReLU}{ReLU}{Rectified Linear Unit}
\newacronym{LR}{LR}{Logistic Regressor}
\newacronym{K-NN}{K-NN}{k-Nearest Neighbours}
\newacronym{GP}{GP}{Gaussian Processes}
\newacronym{SC2}{SC2}{Spectrum Collaboration Challenge}
\newacronym{CIRN}{CIRN}{Collaborative Intelligent Radio Networks}
\newacronym{DU}{DU}{Distributed Unit}
\newacronym{CU}{CU}{Centralized Unit}
\newacronym{RIC}{RIC}{RAN Intelligent Controller}
\newacronym{RT}{RT}{Real-Time}
\newacronym{ATSSS}{ATSSS}{3GPP Access Traffic Steering Switching and Splitting}
\newacronym{RRM}{RRM}{Radio Resource Management}
\newacronym{MANO}{MANO}{Management and Orchestration}
\newacronym{SMO}{SMO}{Service Management and Orchestration}
\newacronym{UE}{UE}{User Equipment}
\newacronym{NFV}{NFV}{Network Function Virtualization}
\newacronym{SDN}{SDN}{Software-Defined Networking}
\newacronym{NIF}{NIF}{Network Intelligence Function}
\newacronym{NIFD}{NIFD}{NIF Descriptor}
\newacronym{NIS}{NIS}{Network Intelligence Service}
\newacronym{NISD}{NISD}{NIS Descriptor}
\newacronym{NIF-C}{NIF-C}{NIF Component}
\newacronym{NS}{NS}{Network Service}
\newacronym{NIP}{NIP}{Network Intelligence Plane}
\newacronym{NIO}{NIO}{Network Intelligence Orchestration}
\newacronym{IO}{IO}{Intelligence Orchestrator}
\newacronym{B5G}{B5G}{Beyond 5G}
\newacronym{MLOps}{MLOps}{ML Operations}
\newacronym{FR}{FR}{Functional Requirements}
\newacronym{NFR}{NFR}{Non-Functional Requirements}
\newacronym{API}{API}{Application Programming Interface}
\newacronym{SDO}{SDO}{Standard-Defining Organization}
\newacronym{OSM}{OSM}{Open Source MANO}
\newacronym{ONAP}{ONAP}{Open Network Automation Platform}
\newacronym{MILP}{MILP}{Mixed-Integer Linear Programming}
\newacronym{NSS}{NSS}{Network Slice State}
\newacronym{N-MAPE-K}{N-MAPE-K}{Network Monitor-Analyze-Plan-Execute over a shared Knowledge}
\newacronym{eMBB}{eMBB}{Enhanced Mobile Broadband}
\newacronym{URLLC}{URLLC}{Ultra-Reliable Low Latency Communications}
\newacronym{NRF}{NRF}{Network Repository Function}
\newacronym{NEF}{NEF}{Network Exposure Function}
\newacronym{NWDAF}{NWDAF}{Network Data Analytics Function}
\newacronym{ETSI}{ETSI}{European Telecommunications Standards Institute}
\newacronym{3GPP}{3GPP}{3rd Generation Partnership Project}
\newacronym{O-RAN}{O-RAN}{Open-Radio Access Network}
\newacronym{5GPPP}{5GPPP}{5G Infrastructure Public Private Partnership}
\newacronym{WG}{WG}{Working Group}
\newacronym{NISt}{NI Stratum}{Network Intelligence Stratum}
\newacronym{IoT}{IoT}{Internet of Things}
\newacronym{RL}{RL}{Reinforcement Learning}
\newacronym{SLA}{SLA}{Service Level Agreement}
\newacronym{O-CU}{O-CU}{Open Central Unit}
\newacronym{O-DU}{O-DU}{Open Distributed Unit}
\newacronym{5GC}{5GC}{5G Core}
\newacronym{OAM}{OAM}{Operations, Administration, and Maintenance}
\newacronym{MDAF}{MDAF}{Management Data Analytics Function}
\newacronym{MDAS}{MDAS}{Management Data Analytics Services}
\newacronym{AF}{AF}{Application Function}
\newacronym{TLS}{TLS}{Transport Layer Security}
\newacronym{CSOI}{CSOI}{Creation Selection Optimization and Instantiation}
\newacronym{PolicyIC}{PolicyIC}{Policy Interpreter and Configuration}
\newacronym{XAI}{XAI}{Explainable AI}
\newacronym{XR}{XR}{eXtended Reality}
\newacronym{ENI}{ENI}{Experiential Networked Intelligence}

\journal{Computer Networks}

\newcommand\copyrighttext{%
  \footnotesize \copyright~2024. This is the author's version of an article submitted to the SI: Evolution of Networked AI Systems: Trends, Challenges, and Opportunities at Computer Networks, Elsevier. This work is licensed under CC BY-NC-ND 4.0. \\
  To view a copy of this license, visit http://creativecommons.org/licenses/by-nc-nd/4.0/}
\newcommand\copyrightnotice{%
\begin{tikzpicture}[remember picture,overlay]
\node[anchor=north,yshift=-10pt] at (current page.north) {\fbox{\parbox{\dimexpr0.9\textwidth-\fboxsep-\fboxrule\relax}{\copyrighttext}}};
\end{tikzpicture}%
}

\begin{document}

\begin{frontmatter}

\title{Designing the Network Intelligence Stratum for 6G Networks}

\author[1,2]{Paola Soto}
\ead{paola.soto-arenas@uantwerpen.be}
\author[1]{Miguel Camelo}
\author[3]{Gines Garcia-Aviles}
\author[3]{Esteban Municio}
\author[4]{Marco Gramaglia}
\author[5]{Evangelos Kosmatos}
\author[1]{Nina Slamnik-Kriještorac}
\author[6]{Danny De Vleeschauwer}
\author[7]{Antonio Bazco-Nogueras}
\author[8]{Lidia Fuentes}
\author[8]{Joaquin Ballesteros}
\author[9]{Andra Lutu}
\author[10]{Luca Cominardi}
\author[10]{Ivan Paez}
\author[7,4]{Sergi Alcalá-Marín}
\author[7]{Livia Elena Chatzieleftheriou}
\author[11]{Andres Garcia-Saavedra}
\author[7]{Marco Fiore}

\affiliation[1]{organization={University of Antwerp - imec, IDLab},
            city={Antwerp}, 
            country={Belgium}
            }
\affiliation[2]{
            organization={Universidad de Antioquia},
            city={Medellín}, 
            country={Colombia}
            }
\affiliation[3]{organization={i2CAT},
            city={Barcelona}, 
            country={Spain}
            }
\affiliation[4]{organization={University Carlos III},
            city={Madrid}, 
            country={Spain}
            }
\affiliation[5]{organization={WINGS ICT Solutions},
            city={Athens}, 
            country={Greece}
            }
\affiliation[6]{organization={Nokia Bell Labs},
            city={Antwerp}, 
            country={Belgium}
            }
\affiliation[7]{organization={IMDEA Networks Institute},
            city={Madrid}, 
            country={Spain}
            }
\affiliation[8]{organization={Universidad de Málaga},
            city={Malaga}, 
            country={Spain}
            }
\affiliation[9]{organization={Telefonica Investigacion y Desarrollo SA},
            city={Madrid}, 
            country={Spain}
            }
\affiliation[10]{organization={ZettaScale Technology},
            city={Saint Aubin}, 
            country={France}
            }
\affiliation[11]{organization={NEC Laboratories Europe GmbH},
            city={Madrid}, 
            country={Spain}
            }


\begin{abstract}
As network complexity escalates, there is an increasing need for more sophisticated methods to manage and operate these networks, focusing on enhancing efficiency, reliability, and security. A wide range of \ac{AI}/\ac{ML} models are being developed in response. These models are pivotal in automating decision-making, conducting predictive analyses, managing networks proactively, enhancing security, and optimizing network performance. They are foundational in shaping the future of networks, collectively forming what is known as \ac{NI}. Prominent \acp{SDO} are integrating \ac{NI} into future network architectures, particularly emphasizing the closed-loop approach. However, existing methods for seamlessly integrating \ac{NI} into network architectures are not yet fully effective. This paper introduces an in-depth architectural design for a \ac{NISt}. This stratum is supported by a novel end-to-end \ac{NI} orchestrator that supports closed-loop \ac{NI} operations across various network domains. The primary goal of this design is to streamline the deployment and coordination of \ac{NI} throughout the entire network infrastructure, tackling issues related to scalability, conflict resolution, and effective data management. We detail exhaustive workflows for managing the \ac{NI} lifecycle and demonstrate a reference implementation of the NI Stratum, focusing on its compatibility and integration with current network systems and open-source platforms such as Kubernetes and Kubeflow, as well as on its validation on real-world environments. The paper also outlines major challenges and open issues in deploying and managing \ac{NI}. 
\end{abstract}

\end{frontmatter}


\copyrightnotice


\section{Introduction}\label{sec:introduction}

To meet the stringent demands imposed by upcoming services such as multisensory \ac{XR} applications and connected robotics~\cite{Saad2020}, which will rely on performance metrics such as virtually unlimited capacity and perceived zero latency, 6G networks will necessitate advanced algorithms. These algorithms will be executed by diverse controllers and orchestrators, leveraging \ac{AI} and \ac{ML} techniques, managing different micro-domains within the network. They will autonomously oversee the intricate assembly of \acp{NF} and associated resources, forming a mosaic to support various \acp{NS}, such as network slices, utilized by different tenants. This orchestration contributes to the development of intelligent networking, referred to as \ac{NI}. More specifically, an instance of \ac{NI} is described as a sequence of efficient \ac{AI}/\ac{ML} algorithms that rapidly identify or predict new requests or fluctuations in network activities~\cite{camelo2022requirements}. Subsequently, these algorithms respond by automatically instantiating, relocating, or re-configuring \acp{VNF}.

Hence, the effectiveness and sustainability of 6G systems will heavily rely on the seamless integration of \ac{NI} solutions into the network infrastructure. Within the network framework, each controller and orchestrator is expected to execute multiple instances of \ac{NI}, aligning with various \ac{KPI} objectives. These objectives encompass guarantees related to \ac{QoS} and \ac{QoE}, infrastructure and resource utilization optimization across diverse tenants or network services, and the realization of end-to-end network automation for achieving zero-touch network and service management. Consequently, the architectural blueprint of mobile networks requires a comprehensive reconsideration, ensuring that the operations of multiple \ac{NI} instances can be seamlessly accommodated across all micro-domains through fully automated processes~\cite{etsi-zsm}.

Ongoing initiatives led by major \acp{SDO} to incorporate \ac{NI} into next-generation network architectures invariably revolve around the concept of ``closed-loop AI''~\cite{wang2020design}. Under this paradigm, \ac{NI} instances deployed at orchestrators and controllers function within closed control loops: they capture the context of management decisions, gather feedback on decision quality through continuous monitoring, and use it to enhance future decision-making. The closed-loop model enables \ac{NI} to comprehend the significance of specific factors in a given situation and progressively automate decision-making in alignment with targeted \acp{KPI}.

However, existing frameworks for network management established by prominent SDOs such as \ac{3GPP} and \ac{ETSI}, along with global industrial initiatives such as \ac{O-RAN}, currently fall short in facilitating the seamless integration of closed-loop \ac{NI}. This hinders the practical adoption of \ac{NI} within 6G networks and calls for original enhancements to the network architectural paradigm. In particular, it is fundamental that a native integration of \ac{NI} algorithms into the overall mobile network architecture is considered to fully support automation capabilities in future generations of communication systems.

This paper presents the complete architectural design and supporting procedures to realize a \ac{NISt} for 6G networks. Specifically, the contributions of this paper are fourfold:

\begin{itemize}
    \item A complete architectural design of a \ac{NISt}, including the \ac{NIO} and the tools to define and design \ac{NI} in a unified manner. This paper integrates the findings and results of our preliminary work~\cite{camelo2022requirements, camelo2022daemon, gramaglia2022network, chatzieleftheriou2023orchestration}, presenting them in a condensed and cohesive format. In addition, compared to state-of-the-art works (see Table~\ref{tab:related-work}), our \ac{NISt} tackles simultaneously multiple challenges of \ac{NI} such as how to properly define \ac{NI}, how to manage their life-cycle, and how to coordinate them via well-defined procedures. 
    
    \item We define a set of internal interfaces to facilitate the interaction among functionalities in the \ac{NISt}. Moreover, we analyze how the external interfaces may be realized when interacting with the \ac{RAN} and Core network segments. To the authors' best knowledge, this is the first work that provides such interfaces.
    
    \item We provide a detailed description of several inter- and intra-\ac{NIO} procedures that are required to realize the \ac{NISt}. Compared to our previous work~\cite{chatzieleftheriou2023orchestration}, we not only define them as functionalities but also provide the required workflows and sequence diagrams that allow us to perform these procedures. 
    
    \item We implement and demonstrate~\cite{daemon_demo_nip} some of the capabilities of the \ac{NISt} via two use cases. While the orchestration of \ac{NIS} with support of \ac{ML} pipelines was part of our previous work~\cite{chatzieleftheriou2023orchestration}, we include a new implementation where we show the capabilities of the \ac{NISt} to orchestrate two \ac{NIF} and realize a \ac{NIS} in the context of service orchestration at the edge.  
\end{itemize}

The remainder of this paper is organized as follows. Section~\ref{sec:related} reviews and summarizes the research and standardization efforts in defining an \ac{AI}-native architecture. Section~\ref{sec:architecture} shows the complete design of the \ac{NISt}, including our methodology for defining \ac{NI} and detailing the necessary functional blocks to orchestrate \ac{NI} in an end-to-end way. Moreover, Section~\ref{sec:interfaces} presents internal and external interfaces needed to perform such orchestration. Using the functional blocks and the interfaces defined in the previous sections, Section~\ref{sec:procedures} shows their interactions to perform the most important orchestration tasks (e.g., creation, management, and termination), including the identified challenges regarding knowledge sharing and conflict resolution. Finally, Section~\ref{sec:results} shows a reference implementation of the proposed \ac{NISt}. Section~\ref{sec:conclusion} concludes the paper and gives an overview of the future challenges. 

\section{Research and standardization efforts towards the integration of NI in mobile networks}\label{sec:related}

The vision of \ac{B5G} and 6G networks largely builds on conflict-free and synergic operation among various \ac{NI} algorithms across network schedulers, controllers, and orchestrators~\cite{camelo2022requirements,gramaglia2022network}. This section reviews current efforts from \acp{SDO} and academia that aim to facilitate joint and end-to-end \ac{NI} operation. 

Current network management frameworks proposed by major \acp{SDO} present deficiencies when integrating and managing \ac{NI} approaches. Our examination, detailed in~\cite{camelo2022daemon} and in Section~5 and Appendix~B of~\cite{wp2_d21}, reveals that standards and platforms from entities like \ac{ETSI}, \ac{O-RAN}, and \ac{3GPP}, including implementations such as \ac{OSM} or \ac{ONAP}, lack mechanisms for coordinating intelligence across different network micro-domains and providing solutions for decentralized and unified data management across \ac{NI} instances. Moreover, these frameworks show minimal support for managing the \ac{NI} lifecycle (e.g., \ac{O-RAN}) and only early consideration for methodologies defining and representing \ac{NI} models (e.g., \ac{ETSI}-\ac{ENI}). Table~\ref{tab:related-work} summarizes the existing frameworks' functionalities for \ac{NI} management in end-to-end control and orchestration of networks.

Noticing that current network management frameworks proposed by major \acp{SDO} are not yet \ac{AI}-native, researchers have delved into various aspects, including the development of \ac{NI} algorithms, the orchestration of \ac{NI} instances, and the coordination of \ac{NI} across diverse network domains. Investigations into the challenges, methodologies, and advancements in integrating \ac{NI} aim to pave the way for more efficient, scalable, and autonomous network operations. By examining prior works, this paper positions its contributions within the broader context of ongoing endeavors to enhance the intelligence and adaptability of network architectures through the seamless integration of the so-called \acrlong{NI}.

Ericsson, a major player in the telecom industry, addresses the escalating use of \ac{AI} in networks in its whitepaper~\cite{ericsson2023ainative}. Their definition of ``\ac{AI}-native'' emphasizes the comprehensive integration of \ac{AI}, which requires a corresponding data infrastructure in every sub-component of an entity, as opposed to adding \ac{AI} components to non-\ac{AI}-based entities. This approach extends across multiple layers and domains, with a crucial requirement being a distributed data infrastructure supporting (re-)training of models. Ericsson also introduces an \ac{AI}-native maturity model as a tool for assessing a product's position on the \ac{AI}-native spectrum and planning the evolution of its implementation towards \ac{AI}-native capabilities. Additionally, they emphasize the importance of \ac{MLOps} functions for comprehensive end-to-end model lifecycle management within an \ac{AI}-native architecture.

\begin{table*}
\centering
\caption{Main differences between our work and similar approaches proposed by major SDOs and academia.}
\label{tab:related-work}
\resizebox{\textwidth}{!}{%
\begin{tabular}{|c|c|c|c|c|c|}
\hline
\textbf{\begin{tabular}[c]{@{}c@{}}Framework /\\ Related Work\end{tabular}}              & \textbf{\begin{tabular}[c]{@{}c@{}}Methodology \\ to define NI\end{tabular}} & \textbf{\begin{tabular}[c]{@{}c@{}}Mechanisms to manage \\ the lifecycle of NI\end{tabular}} & \textbf{\begin{tabular}[c]{@{}c@{}}Mechanisms to coordinate \\ NI across different \\ network segments\end{tabular}} & \textbf{\begin{tabular}[c]{@{}c@{}}Decentralized and\\ unified data management\\ for NI instances\end{tabular}} & \textbf{\begin{tabular}[c]{@{}c@{}}Mechanisms to\\ solve conflicts\end{tabular}} \\ \hline
ETSI MEC                                                                                 & No                                                                           & No                                                                                           & No                                                                                                                   & No                                                                                                              & No                                                                               \\ \hline
ETSI NFV                                                                                 & No                                                                           & No                                                                                           & No                                                                                                                   & No                                                                                                              & No                                                                               \\ \hline
ETSI ENI                                                                                 & Yes                                                                          & No                                                                                           & No                                                                                                                   & No                                                                                                              & No                                                                               \\ \hline
O-RAN                                                                                    & Yes                                                                          & Partially                                                                                    & No                                                                                                                   & No                                                                                                              & No                                                                               \\ \hline
Open Source MANO (OSM)                                                                   & No                                                                           & No                                                                                           & No                                                                                                                   & No                                                                                                              & No                                                                               \\ \hline
3GPP                                                                                     & No                                                                           & No                                                                                           & No                                                                                                                   & No                                                                                                              & No                                                                               \\ \hline
ONAP                                                                                     & No                                                                           & No                                                                                           & No                                                                                                                   & No                                                                                                              & No                                                                               \\ \hline
\begin{tabular}[c]{@{}c@{}}Ericssson Whitepaper~\cite{ericsson2023ainative}\end{tabular}                  & No                                                                           & Partially                                                                                    & Partially                                                                                                            & Partially                                                                                                       & No                                                                               \\ \hline
\begin{tabular}[c]{@{}c@{}}Li et al.~\cite{li2022distributed}\end{tabular}                            & No                                                                           & Partially                                                                                    & Partially                                                                                                            & Partially                                                                                                       & Partially                                                                        \\ \hline
\begin{tabular}[c]{@{}c@{}}Rossi et al.~\cite{rossi2022network}\end{tabular}                         & Partially                                                                           & No                                                                                           & Partially                                                                                                            & Partially                                                                                                       & No                                                                               \\ \hline
\begin{tabular}[c]{@{}c@{}}Brito et al.~\cite{brito2023network}\end{tabular}                         & No                                                                           & Yes                                                                                          & Yes                                                                                                                  & Yes                                                                                                             & No                                                                               \\ \hline
\begin{tabular}[c]{@{}c@{}}D'oro et al. -- OrchestRAN~\cite{d2023orchestran}\end{tabular}                    & No                                                                           & Partially                                                                                    & Partially                                                                                                            & No                                                                                                              & Yes                                                                              \\ \hline
\textbf{\begin{tabular}[c]{@{}c@{}}Network Intelligence Stratum\\ (Our Work)\end{tabular}} & \textbf{\begin{tabular}[c]{@{}c@{}}Yes~\cite{camelo2022requirements, camelo2022daemon}\end{tabular}}               & \textbf{\begin{tabular}[c]{@{}c@{}}Yes\\ (This work)\end{tabular}}                           & \textbf{\begin{tabular}[c]{@{}c@{}}Yes~\cite{camelo2022daemon, chatzieleftheriou2023orchestration}\\ (Extended in this work)\end{tabular}}                             & \textbf{\begin{tabular}[c]{@{}c@{}}Yes~\cite{camelo2022requirements}\end{tabular}}                                                  & \textbf{\begin{tabular}[c]{@{}c@{}}Yes\\ (This work)\end{tabular}}               \\ \hline
\end{tabular}%
}
\end{table*}

In~\cite{li2022distributed}, Li et al. explored native intelligence solutions for 6G networks, delving into the distributed network architecture of native intelligence. It highlights the capabilities of individual intelligent nodes and the importance of collaboration among them. The discussion extends to cross-domain coordination and knowledge sharing and suggests potential solutions, emphasizing the combination of distributed learning and network functionalities. More than proposing an \ac{AI}-native architecture, the paper lists the requirements that \ac{AI}-based functionalities pose over the current network architecture and explains how it should evolve towards an \ac{AI}-native architecture. 

Rossi et al.~\cite{rossi2022network} present a vision for future networks wherein \ac{AI} attains the status of a primary commodity. The foundational principle revolves around ``fast and slow" types of \ac{AI} reasoning, each offering distinct capabilities for processing network data. Similar to the different timescales in data processing, they propose that intelligence should also operate in different timescales. Fast intelligence is used for perception tasks, where the bias in the \ac{NI} models is not noticeable or can be tolerated. Instead of proposing a closed-loop control, the authors pose the Data, Information, Knowledge, and Wisdom (DIKW) pyramid as the main building blocks of native network intelligence. In such a pyramid, there are two prominent closed loops: one between the data and the information, where the fast intelligence resides, and another between the information and the knowledge, where a more robust yet slower intelligence oversees and controls the fast intelligence instances. 

Brito et al. suggest a network architecture for implementing \ac{AI}-based applications across various network domains~\cite{brito2023network}. The aim is to prevent the formation of \ac{AI} silos by providing reusable data and models, ensuring scalable deployments. Their \ac{AI}-native architecture is composed of the Network and Service Automation Platform (NSAP) and the Connect-Compute platform (CCP). The NSAP spans multiple network domains and is responsible for optimizing and managing \acp{NIF}. On the other hand, the CCP ensures the necessary lifecycle operations for each \ac{NIF}. Besides delineating the architecture, the authors furnish workflows for the comprehensive management of \ac{AI}-based applications and demonstrate the feasibility of the architecture through a vehicular use case.

Focusing on the architectural models proposed by the \ac{O-RAN} Alliance, D'Oro et al. devised \textit{OrchestRAN}, a \ac{NI} orchestration framework for next-generation systems~\cite{d2023orchestran}. Specifically, the authors showed that the intelligence orchestration problem is NP-hard and proposed three complexity reduction techniques. Deployed as a rApp in the non-\ac{RT} \ac{RIC}, \textit{OrchestRAN} empowers Network Operators (NOs) to define high-level control and inference objectives. \textit{OrchestRAN} autonomously determines the optimal set of data-driven algorithms and their execution location, either the cloud or the edge. In this way, the framework can fulfill the intentions specified by NOs, ensuring compliance with desired timing requirements and preventing conflicts between different data-driven algorithms that govern the same parameter set.

\subsection{Differences with previous works}
In the preceding sections, we examined the ongoing endeavors of prominent standardization bodies and academic institutions in formulating an \ac{AI}-native architecture. This subsection aims to underscore the distinctions between our work and these existing initiatives. Specifically, we analyze crucial aspects currently overlooked by present management frameworks and briefly elaborate on how our research addresses these gaps. Table~\ref{tab:related-work} summarizes the main differences between our work and the reviewed literature. 

One key differentiator between the proposed architecture and the related work is how we define \ac{NI}. Most works assume the \ac{NI} instance as a single block, not a composition of multiple blocks. Thanks to the \ac{MAPE-K} control model, we can decompose the \ac{NI} in atomic elements that can be deployed across the whole infrastructure. This approach allows for the unified representation of \ac{NI} instances independently of their inner models, facilitates the re-use of similar blocks among different \acp{NIS}, and helps identify conflicts as demonstrated in the \ac{vRAN} use case presented in~\cite{gramaglia2022network}. A similar approach is presented in~\cite{rossi2022network}, where their DIKW pyramid resembles the \ac{MAPE-K} but lacks the concept of multiple closed-loop control. 

Regarding the \ac{NI} lifecycle management, Brito et al.~\cite{brito2023network} provided appropriate workflows for \ac{NIF}/\ac{NIS} instantiation and replacement. In this paper, we go a step further by providing the necessary workflows for the complete \ac{NI} lifecycle, from its creation until its termination. Although Li et al.~\cite{li2022distributed} mention the \ac{NI} lifecycle management as a main issue in \ac{AI}-native architectures, they did not discuss how it can be solved using the proposed \ac{AI}-native architecture.   

A common denominator among the research work is the importance of data. In~\cite{ericsson2023ainative}, the pervasive nature of the \ac{NI} is intricately linked to a distributed data infrastructure. The ability to execute and, when necessary, train \ac{AI} models relies on the ubiquitous availability of data and computing resources. Moreover, the data ingestion speed defines the ``fast and slow" intelligence in~\cite{rossi2022network}. Despite its importance, the studied works do not present a decentralized and unified data management framework for the \ac{NI} instance. On the contrary, in~\cite{camelo2022requirements}, we analyzed the challenges and requirements imposed by the distribution and management of data among disaggregated infrastructure and the impact of operating at different timescales for control systems. Moreover, we analyzed the \ac{NI} design concerning three fundamental aspects: data, decision-making, and decision enforcement, which are fundamental aspects of realizing the \ac{NISt}.

The work presented in this paper summarizes our previous work regarding the methodology to define \ac{NI}, the importance of data, and the mechanisms to coordinate \ac{NI} instances. Additionally, it extends our previous work by providing the interfaces that allow the stratum to provide communication between their functionalities (internals) and with external entities. Moreover, we provide a set of detailed procedure workflows for managing the \ac{NI} lifecycle, including detailed procedures to resolve conflicts among multiple and colliding \ac{NI} and knowledge sharing. Moreover, we extend the reference implementation by considering the necessary extensions to support conflict resolution capabilities in top open-source platforms such as Kubernetes~\cite{kubernetes}, Kubeflow~\cite{kubeflow}, and Zenoh~\cite{zenoh}.   

\section{Architectural design of the NI stratum}\label{sec:architecture}
\begin{figure*}
    \centering
    \includegraphics[width=.8\textwidth]{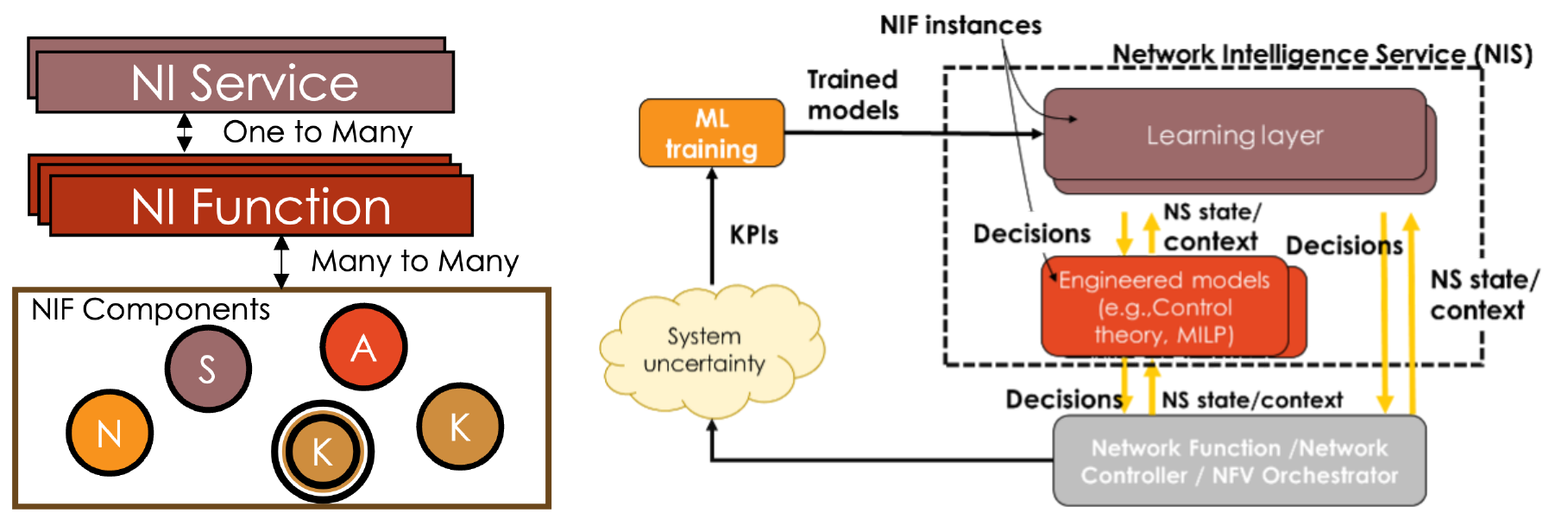}
    \caption{The high-level hierarchical taxonomy of NI algorithms. An NIF corresponds to an individual NI instance that assists a specific functionality; for example, it could capture the implementation of a capacity forecasting task, assisting an NI edge orchestration functionality.}
    \label{fig:fig7}
\end{figure*}

Our proposal revolves around the complete design of a \textit{\ac{NISt}} designed to fulfill multiple objectives within an \ac{AI}-empowered network infrastructure. Firstly, the \ac{NISt} aims to facilitate closed-loop \ac{NI} systematically throughout the entire end-to-end network architecture. Secondly, it seeks to enable the coordination of various \ac{NI} instances deployed across the network, fostering collaboration, exploiting synergies, and effectively managing conflicts. Thirdly, the \ac{NISt} defines essential interfaces that \ac{NI} algorithms can utilize to interact with their respective local environments. In addressing the limitations of existing academic research and industry related to the \ac{NISt}, this framework is conceived as an orthogonal approach where \ac{NI} instances can effectively be integrated into the traditional planes (data, control, and management) for easy adoption in the industry, complementing existing architectures.

\subsection{Defining and designing intelligence in the network}

From an architectural viewpoint, our conceptualization for \ac{NI} management hinges upon similar design principles to those underpinning the management of \acp{NS} in 5G networks. This approach enables the adaptation of familiar concepts to the domain of network intelligence and facilitates the integration of the \ac{NISt} with existing 5G architectural frameworks. Building on this strategy, and in a manner analogous to the information model outlined for network management by entities like \ac{3GPP}, we introduce the notions of \ac{NIF} and \ac{NIS}, defined as follows.

\textbf{\ac{NIF}}: This functional block within an \ac{NI} instance implements decision-making functionality for deployment in a controller, \ac{NFV} orchestrator, or individual \ac{NF}. It features well-defined interfaces and behavior corresponding to an individual \ac{NI} instance that serves a specific functionality.

\textbf{\ac{NIS}}: An assembly of \acp{NIF} with a specific objective, often associated with a particular set of targeted \acp{KPI}.

To facilitate the modeling of any \ac{NI} algorithm, we represent complex \ac{NI} algorithms as a hierarchy of \acp{NIS} that can be broken down into one or more \acp{NIF}. There is a one-to-many relationship between \gls{NIS} and \glspl{NIF}, as the former could be provided by one or more instances of the latter. \glspl{NIF} themselves could be of different kinds: they could be learning models based on, e.g., \glspl{DNN} or Engineered Models, or they could be built upon specific optimization algorithms such as the ones based on control theory or \gls{MILP}. The heterogeneous definition of \acp{NIF} is not limited to complex \ac{AI} models but also encompasses traditional and interpretable models that are not necessarily data-driven. 

The high-level interactions among the building blocks mentioned above are illustrated in Figure~\ref{fig:fig7}. A \ac{NIF} may engage in two primary interactions with the underlying layers (i.e., control plane, user plane, or infrastructure, represented by an \ac{NFV} Orchestrator): it may (i) contribute decisions and (ii) receive information concerning the network and the contextual state of such a configuration. These two interactions inherently implement a close-loop \ac{NI}. On top of this, a \ac{NIS} represents a collaborative effort involving one or more \acp{NIF}, potentially organized in a hierarchical structure. As an example, a \ac{NIS} might consist of a Learning-type \ac{NIF} providing decisions to an engineered model \ac{NIF}, which, in turn, influences the underlying infrastructure.

\begin{figure}
    \centering
    \includegraphics[width=.9\columnwidth]{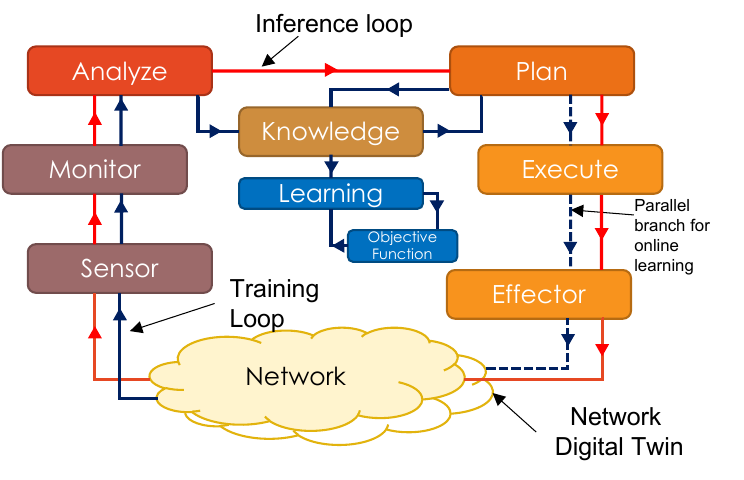}
    \caption{Extended N-MAPE-K abstractions for NI algorithms.}
    \label{fig:fig8}
\end{figure}

When diving into the internal functioning of a single \ac{NIF}, we employ a methodology akin to the \ac{MAPE-K} feedback loop to break down the stages of the closed-loop operation performed by the \acp{NIF}. \ac{MAPE-K} is recognized as one of the most influential reference control models for autonomic and self-adaptive systems~\cite{mapek2006ibm}, yet cannot fully support the specifications of \acp{NIF} internals. Therefore, we introduce an extended \ac{N-MAPE-K}~\cite{camelo2022requirements} model tailored to the \ac{NI} environment, which augments the legacy \ac{MAPE-K} with original training and closed control loops that a \ac{NIF} may implement, as shown in Figure~\ref{fig:fig8}. The \ac{N-MAPE-K} model allows capturing (i) the inference loop, (ii) a traditional supervised training loop, and (iii) a second training loop dedicated to online learning.

Mapping \ac{NI} algorithm components into the \ac{N-MAPE-K} representation allows highlighting the following fundamental classes of atomic \acp{NIF-C}. 

\begin{itemize}
    \item  \textbf{Sensor \acp{NIF-C}} specify all the monitoring probes needed to gather the input measurement data.
    
    \item  \textbf{Monitor \acp{NIF-C}} specify how each \ac{NIF} interacts with the Sensor \acp{NIF-C} and gathers their raw data.
    
    \item  \textbf{Analyze \acp{NIF-C}} include any pre-processing, summary, or data preparation for the specific NI algorithm implemented in the plan \acp{NIF-C}.
    
    \item  \textbf{Plan \acp{NIF-C}} constitute the specific NI algorithm implemented by the \ac{NIF}. 
    
    \item  \textbf{Execute \acp{NIF-C}} specify how the algorithm will interact with the managed system and how to possibly change its configuration parameters.
    
    \item \textbf{Effector \acp{NIF-C}} specify the configuration parameters updated in the \ac{NF}, and the \acp{API} to be used to that end.
\end{itemize}

\subsection{Network Intelligence Stratum}

\begin{figure}
    \centering
    \includegraphics[width=0.90\columnwidth]{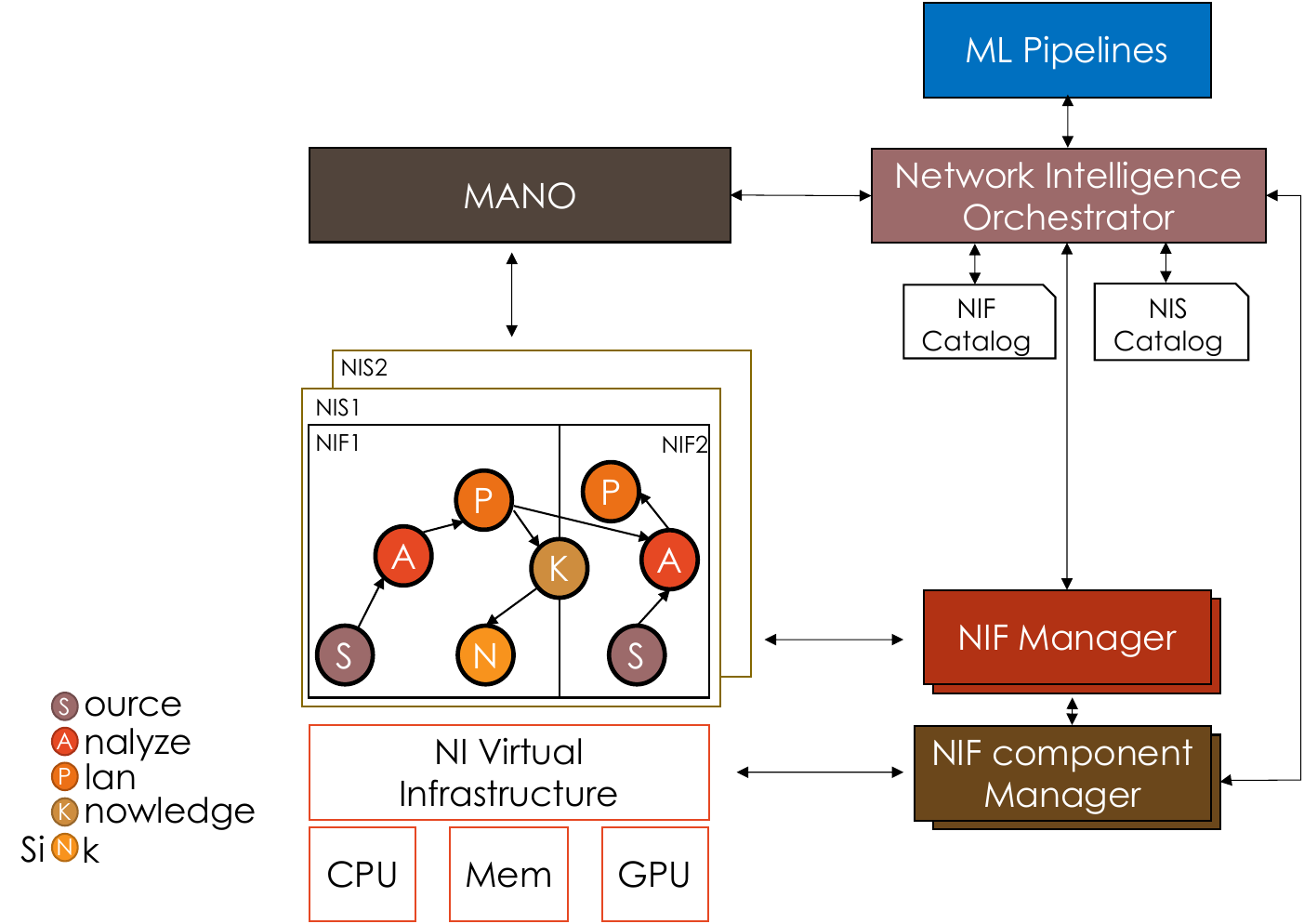}
    \caption{Architecture of the Network Intelligence Stratum.}
    \label{fig:nist}
\end{figure}

In the supervision and coordination of \acp{NIS}, \acp{NIF}, and \acp{NIF-C}, which collectively constitute the overall \ac{NI}, we adapted the layered structure of the \ac{ETSI} \ac{NFV} \ac{MANO} framework. This adaptation tailors the components to the specific requirements of \ac{NI}. The resultant framework forms the \textit{\ac{NISt}} and is illustrated in Figure~\ref{fig:nist}; it is structured into three levels, namely (i) the \ac{NIO}, (ii) the \ac{NIF} Manager, and (iii) the \ac{NIF-C} Manager.

\textbf{\ac{NIF-C} Manager}: This component is responsible for managing the lifecycle of the \ac{NIF-C}. This management encompasses various operations, including onboarding, instantiation, termination, scaling, and state retrieval. The \ac{NIF-C} Manager handles these operations uniformly, regardless of the type of \ac{NIF-C} (i.e., whether it is a \textit{Source}, \textit{Analyze}, \textit{Plan}, \textit{Knowledge}, or \textit{Sink}) and its connection to the network infrastructure. For example, in the case of \textit{Sources}, the IP addresses of different data producers need to be provided, while for \textit{Sink}, specific configuration \ac{API} endpoints must be configured. The instantiation specifics vary based on the context of this interaction. For instance, if the \ac{NIF} operates from the core, \textit{Sinks} and \textit{Sources} integrate with the \ac{NRF} and the \ac{NEF}, synchronizing with the \ac{NWDAF}~\cite{3gpp.23.288}, which captures analytics as a set of \textit{Analyze}, \textit{Plan}, and \textit{Knowledge} components. Similar considerations apply to other network domains, such as the \ac{RAN}, where this framework can be seamlessly integrated with the \ac{O-RAN} xApps or rApps ecosystems~\cite{garcia2021ran}.

\textbf{\ac{NIF} Manager}: The \ac{NIF} Manager, on the other hand, provides a comprehensive overview of the collective \ac{NIF-C} set that forms each \ac{NIF}. In addition to overseeing the lifecycle of the \ac{NIF}, this module is tasked with monitoring the overall health of the intelligence functions. This monitoring involves continuous tracking of learning \acp{KPI} generated by the \acp{NIF}, including metrics like accuracy, particularly when the \ac{NIF} is engaged in inference or serves as an online learning solution. Other metrics, such as loss and training loops, are monitored when the \ac{NIF} is undergoing training. The \ac{NIF} Manager is also responsible for configuring the meta-parameters of the models (via interaction with the \ac{NIF-C} Manager) and conveying the health status of the \ac{NIF} up into the hierarchy to the \ac{NIO}.

\textbf{\acrlong{NIO}}: This module is responsible for overseeing the lifecycle management of the \ac{NIS} by effectively coordinating the \acp{NIF} that constitute each of them. This entails the ability to share \ac{NIF-C} among different \acp{NIF} (e.g., two \acp{NIF} requiring the same input) and establishing arbitration policies when two \acp{NIF} share the same sink, specifically the configuration \acp{API}. Importantly, this coordination occurs at the level of the \ac{NIO}, no longer falling within the purview of the \ac{NIF} Manager. It involves collaboration across \acp{NIF}, necessitating a higher-level perspective uniquely held by the \ac{NIO}. The module also handles connections to network \ac{MANO} frameworks for gathering crucial information, such as expected network \acp{KPI} for the managed slice and service and the status of the underlying network infrastructure. The \ac{NIO} maintains catalogs of already onboarded \ac{NIS} and \acp{NIF}. Notably, \acp{NIF} might require retraining to adapt to changing or diverse conditions, either periodically or on demand. In such cases, the \ac{NIO} interfaces with an external platform to construct \ac{ML} pipelines and execute such operations, exemplified by an \ac{MLOps} framework.

\begin{figure}
    \centering
    \includegraphics[width=\columnwidth]{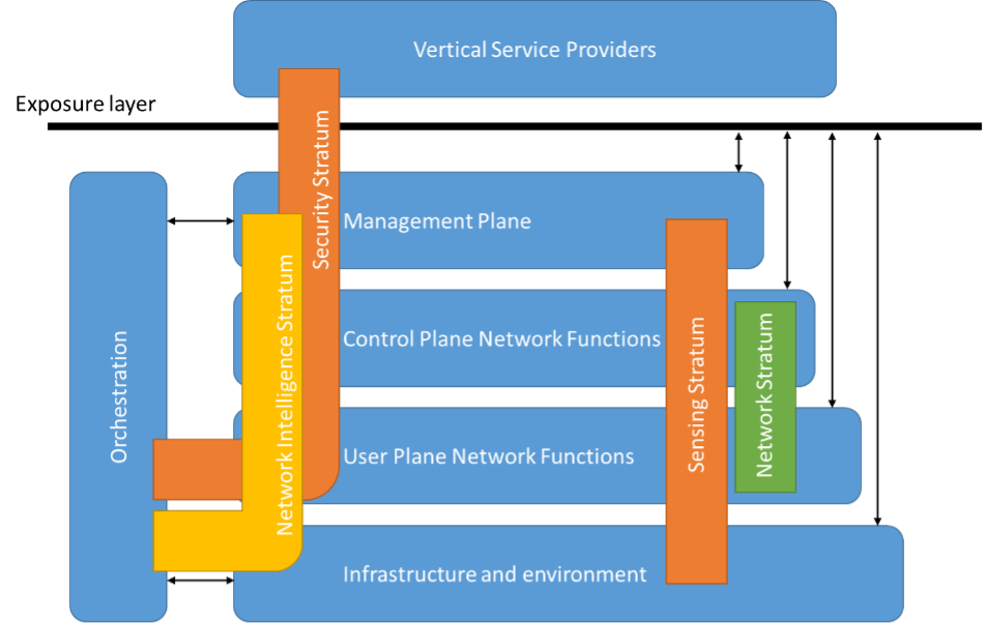}
    \caption{The 5GPPP Architectural WG framework~\cite{bahare_massod_khorsandi_2023_7313232}.}
    \label{fig:fig6}
\end{figure}

The proposed \ac{NISt} moves our previous network intelligence plane design~\cite{camelo2022daemon} from a purely separate plane to a more orthogonal approach where \acp{NIF} and \acp{NIS} can effectively be integrated into the traditional planes (data, control, and management) for easy adoption in the industry.  This term, \textit{Stratum}, has also been embraced as part of the comprehensive architectural framework that has been developed by the \ac{5GPPP} Architecture \ac{WG}, as illustrated in Figure~\ref{fig:fig6} and reported in the whitepaper~\cite{bahare_massod_khorsandi_2023_7313232}, released by the 5G Architecture \ac{WG} in the \ac{5GPPP}. There, the term \textit{Stratum} typically denotes a collection of elements that span various network domains. For example, \textit{network access stratum} encompasses all the elements involved in user registration and authentication across \ac{RAN} and Core. 

\subsection{Functionalities Supported by the NI Stratum}\label{sec:funct-NI-stratum}

The \ac{NISt} is a unified framework that brings together our earlier proposals for (i) the operational hierarchy of \ac{NI} components in the \ac{NISt} and (ii) the \ac{N-MAPE-K} representation of \acp{NIF-C}.  The variety of \acp{NIF} and \acp{NIS} that can be deployed at the network generates new challenges in the way they should be managed that are not presented in current management frameworks. Therefore, in~\cite{chatzieleftheriou2023orchestration}, we discuss the need for specific \ac{NISt} procedures to address challenges arising from the concurrent instantiation of various \acp{NIF} and \acp{NIS}. The challenges are exemplified by using two functionalities to improve the resiliency of a \ac{vRAN} system~\cite{garcia2021nuberu},~\cite[Sec. 2.5]{andres_garcia_saavedra_2022_7525876}. The main challenges to be managed by the \ac{NISt} are Conflict Resolution, Knowledge Sharing among \acp{NIF}, Model Selection, Catalog, and Re-training. Moreover, the \ac{NISt} incorporates functionalities such as Data Analytics, Knowledge Management, Monitoring, \ac{NIS} Lifecycle Management, \ac{NIS} Creation/Selection, Optimization, and Instantiation, Model Explainability, Policy Interpreter and Configuration, \ac{NIS} Workflow Configuration, Network \ac{MANO} Framework, and Conflict Detection and Resolution.

Conflict resolution capabilities are crucial for efficiently reusing and combining elements across \acp{NIF} to build \acp{NIS}. By representing \acp{NIF} as atomic \acp{NIF-C} within the \ac{N-MAPE-K} framework, conflicts may arise in the sharing of different \ac{NIF-C} elements when composing \acp{NIF} to create \acp{NIS}. In the two \ac{NIF} examples mentioned earlier, conflicts may occur when monitoring data, requiring the \ac{NIF} Manager to ensure information arrives with the necessary granularity, and in policy enforcement, where different \ac{NI} algorithms may configure the same network functions differently. The \ac{NIO} is tasked with deploying conflict resolution policies to guarantee optimal decisions, overseeing individual \acp{NIF}, and monitoring access to data sources and policies to amend sub-optimal decisions.

Additionally, the \ac{NIO} plays a crucial role in providing centralized coordination among multiple \acp{NIF}, enabling knowledge sharing for synergistic performance improvements. For example, the knowledge learned by a \ac{NIF} can support other \ac{NIF}'s decisions, and vice versa. Such knowledge-sharing capability can extend accross domains: for instance, in Section 4.2 of~\cite{wp4_d42}, the presence of an anomaly detection solution for \ac{IoT} platforms, where the user plane traverses multiple domains. In this scenario, the \ac{NIO} facilitates synchronization among parties involved in building the user plane for \ac{IoT} devices, addressing challenges in root-cause analysis of anomalies.

Model Selection, Catalog, and Re-training are essential for \ac{NIS} to adapt to the underlying environment. While not directly derived from \ac{NI} algorithms' design, \ac{NIS} may require knowledge of the software/hardware environment and device location. In a pure \ac{ML} environment, tasks are handled by \ac{MLOps} frameworks like Kubeflow~\cite{kubeflow} and MLflow~\cite{mlflow}. However, in an \ac{NI}-native architecture, close interaction with the orchestration environment is necessary. The \ac{NIO} ensures that deployed \acp{NIF} match the specific hardware-software-environmental characteristics of network functions. It exchanges execution context information with the \ac{MANO} system to select the appropriate model for inference within a \ac{NIF}. This involves maintaining a model catalog from which the \ac{NIO} selects the most suitable model based on the network's infrastructural status. If no model is available, the \ac{NIO} can invoke training of a new model, fetching the required data as the target algorithm needs.

\begin{figure*}
    \centering
    \includegraphics[width=0.95\textwidth]{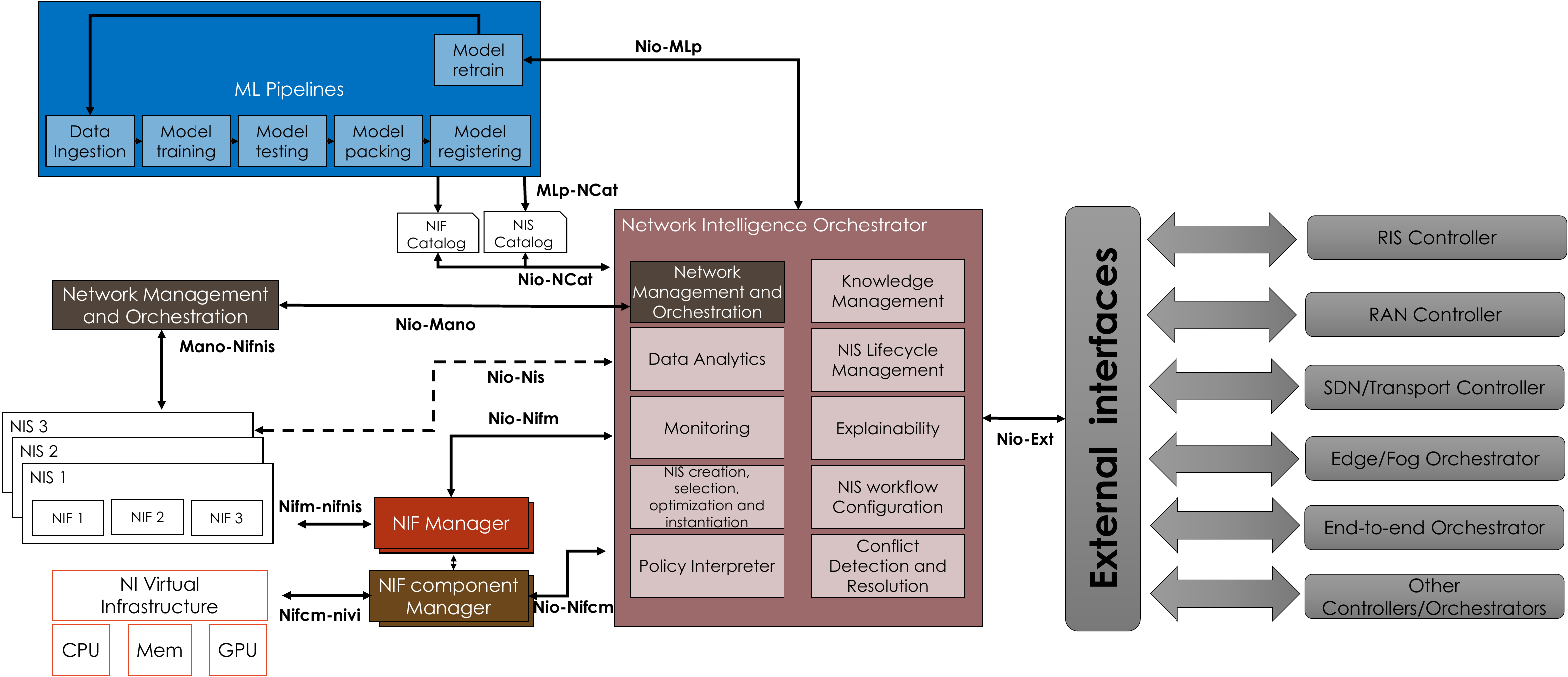}
    \caption{The NI Stratum and the functional blocks of the NIO and ML pipelines, with the internal and external interfaces of the Stratum.}
    \label{fig:fig10-11}
\end{figure*}

The \ac{NIO} plays a central role in managing and coordinating \acp{NIF} to enable \ac{NI}-Native architectures. In response to challenges in deploying multiple \acp{NIF} concurrently, the \ac{NIO} incorporates several key functionalities, which are summarized in Figure~\ref{fig:fig10-11} and listed as follows.

\begin{itemize}
    \item \textit{Data Analytics}: Involves pre-processing or preparing data for the \acp{NIF} by computing statistical measures (e.g., averages, variance, maximum or minimum values) or more complex features (e.g., embeddings via autoencoders or dimensionality reduction techniques, aggregations via clustering algorithms).

    \item \textit{Knowledge Management}: Critical for planning, organizing, acting, and controlling knowledge across all deployed \acp{NIS}.

    \item \textit{Monitoring}: Processes information from \acp{NIS}, covering both \ac{ML}-related (model-specific metrics, data drift) and non-\ac{ML}-related aspects (\ac{QoE}, \ac{QoS}), monitoring \acp{NI} in both training and inference deployments.

    \item \textit{\ac{NIS} Lifecycle Management}: Handles deployment and maintenance of \ac{ML} models, aligning with \ac{MLOps} practices, including the creation of new \ac{ML} pipelines for re-training models.

    \item \textit{\ac{NIS} Creation/Selection, Optimization, and Instantiation}: Involves selecting, optimizing, and instantiating \acp{NIS} based on hardware constraints, with the ability to create a new \ac{NIS} if it is unavailable in the catalog.

    \item \textit{Model Explainability}: Provides methods for human experts to understand black-box \ac{ML} algorithms within the \acp{NIS}, aiding in comprehending decision-making processes.

    \item \textit{Policy Interpreter and Configuration}: Interprets high-level user intent objectives associated with different \ac{NIS}, performing changes in policy as needed.

    \item \textit{\ac{NIS} Workflow Configuration}: Integrates data engineering, \ac{ML}, and DevOps to operationalize deployment, monitoring, and lifecycle management in a modular and flexible way.
    
    \item \textit{Conflict Detection and Resolution}: Provides a mechanism to solve trade-offs arising from conflicting objectives in control and user planes, allowing the \ac{NIO} to compare policies among different \acp{NIS} and perform conflict resolution.
\end{itemize}

The \ac{NIO} interacts with the \ac{MANO} framework, synchronizing network slices, tracking the state and health of network slices and functions, and obtaining real-time information about available resources. This collaboration optimizes network operations, enhances resource utilization, and ensures alignment with vertical service providers' requirements for specific vertical service domains. The \ac{NIO}-\ac{MANO} interaction involves eastbound-westbound interfaces, direct extensions to \ac{MANO} modules, and mappings with \ac{MANO} components such as NFV Orchestrator (NFV-O), Virtual Network Function Manager (VNFM), and Virtual Infrastructure Manager (VIM) in frameworks like \ac{ETSI} \ac{NFV} \ac{MANO}.

The architectural design presented in this section is complemented next in the following ways: (i) in Section~\ref{sec:interfaces}, by presenting and discussing the interfaces that are required to allow communication between internal \ac{NISt} components, and the \ac{NISt} components with external entities such as the \ac{RAN} controller, Core system, and local and end-to-end management systems; and, (ii) in Section~\ref{sec:procedures}, by designing the set of procedures that address the needs and challenges introduced in~\cite{chatzieleftheriou2023orchestration} and that motivate the functionalities mentioned above.

\section{Network Intelligence Stratum Interfaces}\label{sec:interfaces}

As shown in Figure~\ref{fig:fig10-11}, the \ac{NISt} is a composition of different functional blocks that aims for the native integration of \ac{NI} in the network by providing the management and orchestration capabilities for \ac{NIF} and \ac{NIS}. Similar to other well-known frameworks for management and orchestration on specific domains, e.g., \ac{NFV}-\ac{MANO}~\cite{etsi-nfv-mano} and \ac{O-RAN}~\cite{oranwg2aimlworkflow}, the functional blocks of the \ac{NISt} have their own set of internal interfaces. Moreover, external interfaces will allow the \ac{NIO} to communicate with external orchestrators, facilitating efficient resource coordination and orchestration of \ac{NI} across diverse network environments for improved interoperability and scalability. In the following, we will provide a high-level definition of such interfaces and what is expected from them. 

\subsection{Internal Interfaces}
To successfully orchestrate and manage \ac{NI}, it is essential to establish seamless communication and coordination among the various functionalities of the \ac{NISt}. In this subsection, we will outline and elaborate on the specific set of internal interfaces presented in Figure~\ref{fig:fig10-11}. These interfaces are the foundation for enabling effective communication and coordination among the different blocks within the \ac{NISt}, ensuring a harmonized and cohesive \ac{NI} management framework. In the following, we present them according to their functional definition, although from an implementation perspective, they could be provided in a service-based fashion.

\begin{itemize}
    \item 	\textbf{Nio-Nifm}. This interface allows communication between the \ac{NIO} and the \ac{NIF} Manager to effectively manage and orchestrate \ac{NIF} instances within the \ac{NISt} framework. It promotes efficient utilization of network resources, optimized network service delivery, and enhanced scalability and flexibility of virtualized \ac{NIF}. Among life-cycle management, the \ac{NIO} relies on the \ac{NIF} Manager to perform operations related to \ac{NIF} instances, including instantiation, scaling, healing, and termination. Via this interface, the \ac{NIF} Manager can also provide monitoring information about the learning performance (e.g., the loss function when trained), or network performance indicators, and trigger healing actions in case of failures, degradations, or conflicts. Moreover, the \ac{NIO} can gather information related to the status of the \acp{NIF} so it can derive analytics to proactively optimize the \acp{NIF} (e.g., by changing the learning model data feeding speed/timescale to mitigate limitation on available computing resources) or control it (e.g., by adding a new input representation of the data or \ac{ML} model to couple it with other \ac{NIF} when instantiating a new \ac{NIS}). Finally, the \ac{NIO} can also gather information from the \acp{NIF} related to explainable capabilities and use it to take better orchestration and coordination actions among \acp{NIF}. Finally, this interface will allow the \ac{NIO} to perform \ac{ML} workload management.
    
    \item 	\textbf{Nio-Nifcm}. This interface allows the \ac{NIO} to request the \ac{NIF-C} for the allocation, placement, and lifecycle management of virtualized infrastructure resources. These resources include computing (GPU, FPGA, CPU, memory), storage, and networking components required to host and run \ac{NIF} instances. It will also allow for gathering information about the utilization and performance of virtualized infrastructure resources. This includes monitoring the allocated resources' availability, capacity, and performance metrics and providing visibility into resource usage and potential bottlenecks. In case of the need for infrastructure policy enforcement, this interface allows the \ac{NIO} to enforce policies and constraints on the virtualized infrastructure resources such as security policies, learning, and \ac{QoE}/\ac{QoS} requirements, or specific compliance regulations that need to be applied to the infrastructure hosting the \acp{NIF} (e.g., data privacy, data anonymity, model isolation or federation, etc.). 
    
    \item 	\textbf{Nifm-Nifnis}. This interface enables the \ac{NIF} Manager to manage the lifecycle of \ac{NIF} instances. It allows the \ac{NIF} manager to perform operations such as \ac{NIF} instantiation, scaling, healing, termination, and update. In the case of configuration and monitoring, this interface allows the \ac{NIF} Manager to provide configuration parameters and policies to the \ac{NIF} through the interface. Additionally, it can collect monitoring data and performance metrics from the \ac{NIF} instances to ensure their proper functioning and adherence to \acp{SLA} in terms of both networking (e.g., \ac{QoS} and \ac{QoE}) and learning (e.g., accuracy). This \ac{NIF} Manager can also perform fault and performance management. The \ac{NIF} Manager receives fault notifications and performance data from the \acp{NIF} through the interface, allowing it to detect and handle any issues that may arise based on policies defined by the \ac{NIO}. This includes fault localization, resolution, performance optimization, and ensuring the desired performance of the \ac{NIF}. Finally, the \ac{NIF} Manager can manage the state and context of the \ac{NIF} instances. It allows the \ac{NIF} Manager to retrieve and update the state information of the \acp{NIF}, including their operational status, configuration parameters, and runtime data. This information is crucial for maintaining the consistency and continuity of the \ac{NIF} operations. This interface can also provide the capabilities to monitor, manage, and orchestrate \ac{NIS} based on abstract data information such as model knowledge (e.g., \ac{NN} weights, expert knowledge encapsulated in rule-based systems) and explainable model data. Moreover, it will gather information about the \ac{NIF} composition to detect possible conflicts in \ac{NIS} before deployment, given its topological structure, or after re-orchestration of the \ac{NIS} when \ac{NIF} are added/removed/changed. In conjunction with the \textit{Nio-Nifm} interface, this interface allows the \ac{NIO} to also configure the \ac{NIS} (e.g., adding a new \ac{NIF} in the \ac{NIS}). In some implementations, the interaction between \ac{NIO} and \ac{NIS} can be done via a specific interface, e.g., a Nio-Nis interface. 
    
    \item 	\textbf{Nifcm-Nivi}. This interface allows the \acp{NIF} to interact with the \ac{NI} virtualized infrastructure, which includes virtual machines, containers, resources for storage, and networking components. This interface allows \acp{NIF} to utilize the underlying infrastructure to perform their designated functions efficiently. For example, allowing an \ac{ML} model to switch among different computing hardware (e.g., CPU, GPU, TPU, or FPGA) and modes (training versus inference). 
    
    \item 	\textbf{Nio-MLp}. This interface enables the \ac{NIO} to enact \ac{ML} model (re-)training via MLOps.
    
    \item   \textbf{MLp-Ncat}. Via this interface, the \ac{ML} pipeline framework in the \ac{NISt} can access the model register, which serves as a critical connection point in managing and organizing \ac{ML} models empowering \ac{NIF}/\ac{NIS} within the pipeline framework. This interface enables seamless integration and coordination between the pipeline framework and the model register, facilitating efficient model versioning, storage, retrieval, and tracking. This interface streamlines the integration of \ac{ML} models within the pipeline, enabling seamless collaboration, reusability, and scalability of models across the \ac{ML} workflow.
    
    \item 	\textbf{Nio-Ncat}. This interface allows the \ac{NIO} to access the catalog of \ac{NIF}/\ac{NIS} available to deploy in the network. By accessing the catalog, the \ac{NIO} can effectively discover, select, compose, onboard, and manage the lifecycle of \ac{NIF}/\ac{NIS} within the \ac{NISt}. The interface enhances the agility, flexibility, and automation capabilities of the \ac{NI} orchestration system, enabling seamless deployment and efficient management of \ac{NIF}/\ac{NIS} within the \ac{NI} virtual infrastructure.
    
    \item	\textbf{Nio-Mano}. The implementation and deployment of the \ac{NIO} can determine whether \ac{MANO} functionalities are integrated within the \ac{NIO} or external to it. This decision primarily involves a trade-off between a self-contained orchestrator capable of creating, instantiating, and deploying legacy \ac{NF}/\ac{NS}, \ac{ML}-only \ac{NIF}/\ac{NIS}, and hybrids \ac{NIF}/\ac{NIS}, and a lighter orchestrator that relies on external \ac{MANO} for tasks such as managing legacy \ac{NIF}/\ac{NIS} as legacy \ac{NF}/\ac{NS}. The Nio-Mano interface is used in the later case. When the \ac{MANO} is deployed as an external functional block of the \ac{NIO} (e.g., in legacy systems where \ac{MANO} functionalities are already in place), this interface provides the communication mechanism to exchange real-time information to track network slices, function states, and resource availability. This synchronization allows the \ac{NIO} to dynamically adapt decisions and efficiently allocate resources based on the current network characteristics. By maintaining an up-to-date view of available resources, including computing power, storage, and network capabilities, the \ac{MANO} can orchestrate network resources effectively and optimize resource utilization, thereby improving performance. Thanks to this interface, the \ac{NIO} can be aware of such optimization. 

    \item	\textbf{Mano-Nifnis}. This interface allows \ac{MANO} functionality (either internal or external to the \ac{NIO}) to perform orchestration commands directly on \ac{NIF}/\ac{NIS}. For example, operations such as deployment, scaling, updating, or decommissioning of \ac{ML} blocks can be performed via this interface, or monitoring and reporting of \ac{ML} metrics for the data analytics and monitoring block. Also advance functions such as security, e.g., to enforce security policies against adversarial attacks on \ac{ML} models, or conflict detection after deployment can interact with the \ac{NIF}/\ac{NIS} via this interface. 
    
    \item 	\textbf{Nio-Ext}. This interface communicates between the \ac{NIO} and external orchestrators/controllers in the network in the same or across multiple domains. This interface will be detailed in the following section.  
\end{itemize}

To promote industry deployment, validation, and widespread adoption of standardized \acp{API}, we highly recommend that these interfaces are designed following an OpenAPI representation in YAML and JSON  where available (e.g., via ETSI or IEEE), similar to the \ac{NFV}-\ac{MANO} core \acp{API}. Moreover, tools to navigate the specifications and report bugs should also be provided to enhance the usability and effectiveness of the OpenAPI representation. 

\subsection{External Interfaces}
The \textbf{Nio-Ext} interface will allow the \ac{NIO} to communicate with external orchestrators/controllers to achieve efficient collaboration, resource coordination, and \ac{NIF}/\ac{NIS} orchestration across heterogeneous network environments (far edge, edge, \ac{RAN}, transport, core, cloud, etc.). The interface enhances interoperability, scalability, and flexibility, effectively managing and orchestrating resources and \ac{NIF}/\ac{NIS} in complex network ecosystems. 

This interface allows for efficient coordination of resources by exchanging information about available resources and their utilization across different domains. This promotes optimal resource allocation and utilization. Secondly, the interface enables collaboration in \ac{NIF}/\ac{NIS} deployments across multiple domains by facilitating the exchange of \ac{NIF}/\ac{NIS}-level information and dependencies between the \ac{NIO} and external orchestrators/controllers. This enables the instantiation, management, and scaling of complex \ac{NIS} across multi-domain and heterogeneous environments. 

Additionally, the interface supports policy management by facilitating the exchange of policy information between the \ac{NIO} and external orchestrators/controllers. This ensures consistent policy implementation and governance across different domain systems. Moreover, the interface enables the exchange of event and alarm information, allowing for proactive event handling, correlation, and remediation across domains. Finally, the interface facilitates information exchange and federation by enabling the sharing of network topologies, hardware capabilities, \ac{NIF}/\ac{NIS} catalogs, and other relevant data (e.g., monitoring information, model weights, etc.), improving decision-making and coordination capabilities among different orchestration systems. In this section, we will describe two specific cases of such interfaces. 

\subsubsection{O-RAN}
The \ac{O-RAN} Alliance is a global community of mobile network operators, vendors, and research institutions established in February 2018. Its primary goal is to drive the development of open, intelligent, and interoperable \ac{RAN} technologies. Founded by AT\&T, Orange, Deutsche Telekom, Docomo, and China Mobile, \ac{O-RAN} is now supported by over 300 organizations, including major operators and vendors. Analysts predict that open v\ac{RAN}s could surpass the conventional \ac{RAN} market by 2028,  generating revenues close to \$20 billion. 

The \ac{O-RAN} architecture is a new approach to building mobile networks that aims to increase flexibility, interoperability, and innovation. It is designed to enable multi-vendor deployments, reduce costs, and improve network performance. Key aspects of the \ac{O-RAN} architecture are presented in~\cite{polese2023understanding}, where a very important aspect of \ac{O-RAN} is the integration of \ac{AI}/\ac{ML} workflows, i.e., \ac{NI} that may be managed by the \ac{NISt}, with the following principles~\cite{oranwg2aimlworkflow}:

\begin{itemize}
    \item 	\textbf{Offline Learning}: In \ac{O-RAN}, even for \ac{RL} scenarios, some amount of offline learning (where a model is trained with offline data before deployment) is always recommended.
    
    \item 	\textbf{Pre-training and Testing}: Any model deployed within the network needs to be trained and tested beforehand. No completely untrained model should be deployed in the network.
    
    \item 	\textbf{Modularity in \ac{ML} Applications}: As a best practice, \ac{ML} applications should be designed in a modular fashion, with the capability to share data without knowledge of each other’s data requirements. The location or nature of a data source should not bind them.
    
    \item 	\textbf{Service Provider’s Deployment Choice}: The criteria for determining where an \ac{ML} application should be deployed (Non-\ac{RT} \ac{RIC} or Near-\ac{RT} \ac{RIC}) may vary between service providers. Therefore, the service provider should decide the deployment scenario for a given \ac{ML} application.
    
    \item 	\textbf{Optimization of \ac{ML} Model for Efficiency and Performance}: To improve execution efficiency and inference performance, the \ac{ML} model should be optimized and compiled considering the hardware capabilities of the inference host. There should be a balance between efficiency and inference accuracy, with acceptable accuracy loss as one of the optimization goals. The optimization parameters should be determined based on this threshold.
\end{itemize}

\begin{figure*}
    \centering
    \includegraphics[width=0.8\textwidth]{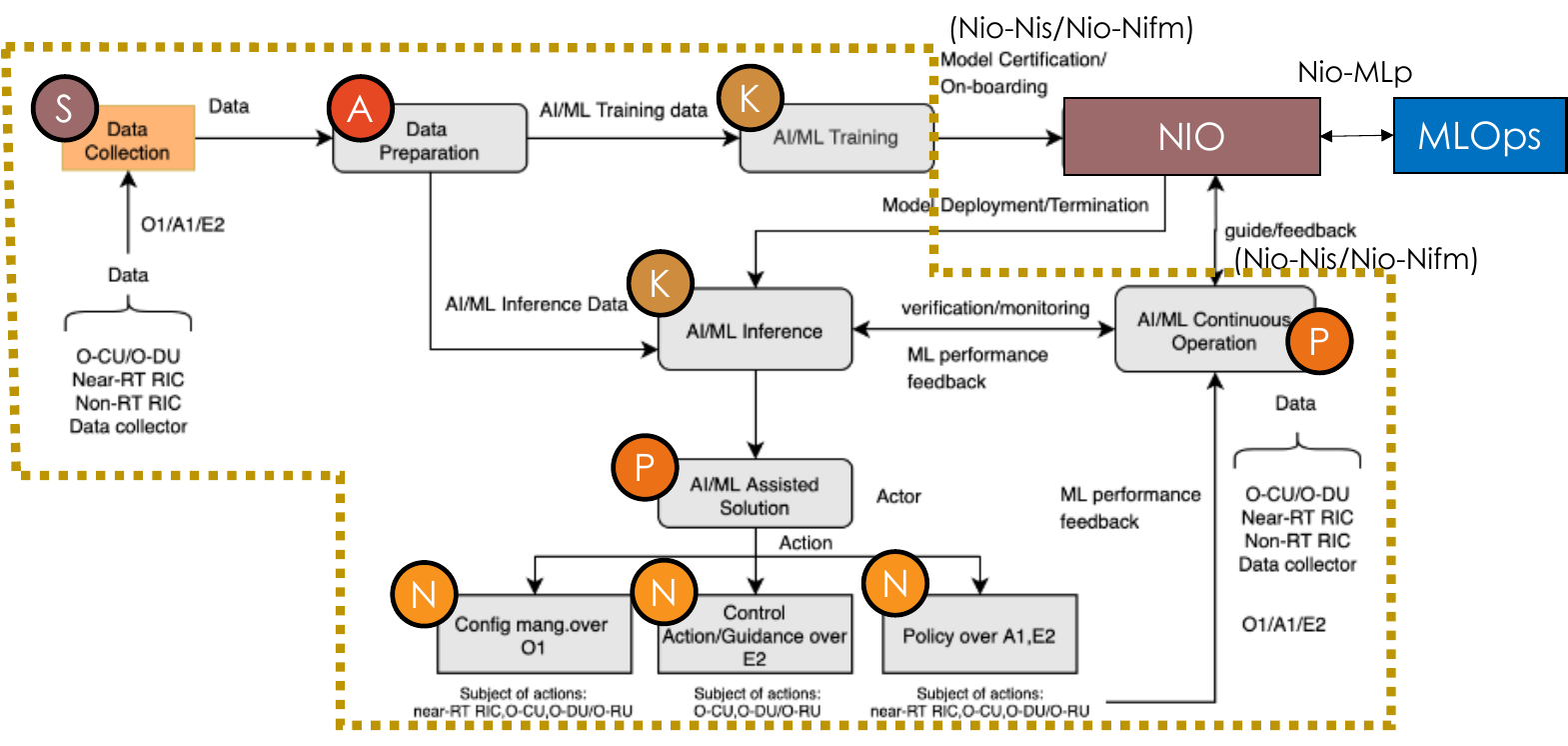}
    \caption{Integration of the NI Stratum and O-RAN AI/ML Lifecycle Procedures and Interface Frameworks}
    \label{fig:fig12}
\end{figure*}

Figure~\ref{fig:fig12} illustrates the general framework of \ac{AI}/\ac{ML} procedures and interfaces and its integration into the proposed \ac{NISt}, including the potential mapping between \ac{ML} components and \ac{O-RAN} components.

\ac{O-RAN} suggests several \ac{AI}/\ac{ML} deployment scenarios that are relevant to our \ac{NISt}; they are summarized as follows:

\begin{itemize}
    \item 	\textbf{Deployment Scenario 1.1}: In this case, \ac{AI}/\ac{ML} Continuous Operation, Model Management, Data Preparation,  Training, and  Inference all take place within the Non-\ac{RT} \ac{RIC} (Non-Real-Time Radio Intelligent Controller).
    
    \item 	\textbf{Deployment Scenario 1.2}: Here, \ac{AI}/\ac{ML} Continuous Operation, Data Preparation for training, and \ac{AI}/\ac{ML} Training are located in non-\ac{RT} \ac{RIC}. However, \ac{AI}/\ac{ML} Model Management is outside non-\ac{RT} \ac{RIC} (either within or outside the \ac{SMO}). Data Collection for inference, Data Preparation for inference, and \ac{AI}/\ac{ML} Inference are in the Near-\ac{RT} \ac{RIC}.
    
    \item 	\textbf{Deployment Scenario 1.3}: \ac{AI}/\ac{ML} Continuous Operation and \ac{AI}/\ac{ML} Inference are within non-\ac{RT} \ac{RIC}. Data Preparation, \ac{AI}/\ac{ML} Training, and Model Management are located outside the non-\ac{RT} \ac{RIC} (either within or outside \ac{SMO}).
    
    \item 	\textbf{Deployment Scenario 1.4}: In this scenario, the non-\ac{RT} \ac{RIC} acts as the \ac{ML} training host for offline model training, and the Near-\ac{RT} \ac{RIC} acts as the \ac{ML} training host for online learning and also as the \ac{ML} inference host.
    
    \item 	\textbf{Deployment Scenario 1.5}: Continuous Operation, Model Management, Data Preparation, and \ac{ML} Training Host are in non-\ac{RT} \ac{RIC}. However, the \ac{O-CU}/\ac{O-DU} acts as the \ac{ML} inference host.
\end{itemize}

Please note that the deployment of ``\ac{AI}/\ac{ML} Continuous Operation" outside of non-\ac{RT} \ac{RIC} is still under study.

\subsubsection{5G-Core}
The \ac{5GC} is one of the most important domains in a \ac{3GPP} mobile system, hence we analyze how the proposed \ac{NISt} can interact with it.  The imperative of network automation drove the design of the \ac{3GPP} system in R15, marking a significant departure from previous releases. In earlier iterations, data generation and analytics in the network primarily relied on proprietary interfaces for exchanges between network elements and their respective managers. However, with R15 and subsequent consolidations, the architecture underwent a comprehensive overhaul to incorporate native support for collecting analytics. As explained below, these analytics can be effectively utilized to establish feedback loops through standardized or proprietary solutions. At the heart of this system lies the \ac{NWDAF}, which performs three key functions: (i) aggregating data, encompassing metrics that reflect the current state of the network, sourced from another producer \acp{NF}; (ii) conducting analytics, involving the computation of refined statistics based on the gathered data; (iii) sharing the computed analytics with other consumer functions across the network.

The generated analytic reports serve as outputs that either present statistics based on historical data or provide predictions for specific metrics, depending on whether the requested timeframe is in the past or future, respectively. These outputs are crucial in optimizing the operation of \acp{NF}. Additionally, the output may include a confidence parameter, ranging from \num{0} to \num{100}, which conveys information about the reliability of the prediction made. Factors determining this confidence parameter may include the volume of data utilized in generating the prediction, the age of the \ac{AI} model employed, and other relevant considerations.

\begin{figure}
    \centering
    \includegraphics[width=\columnwidth]{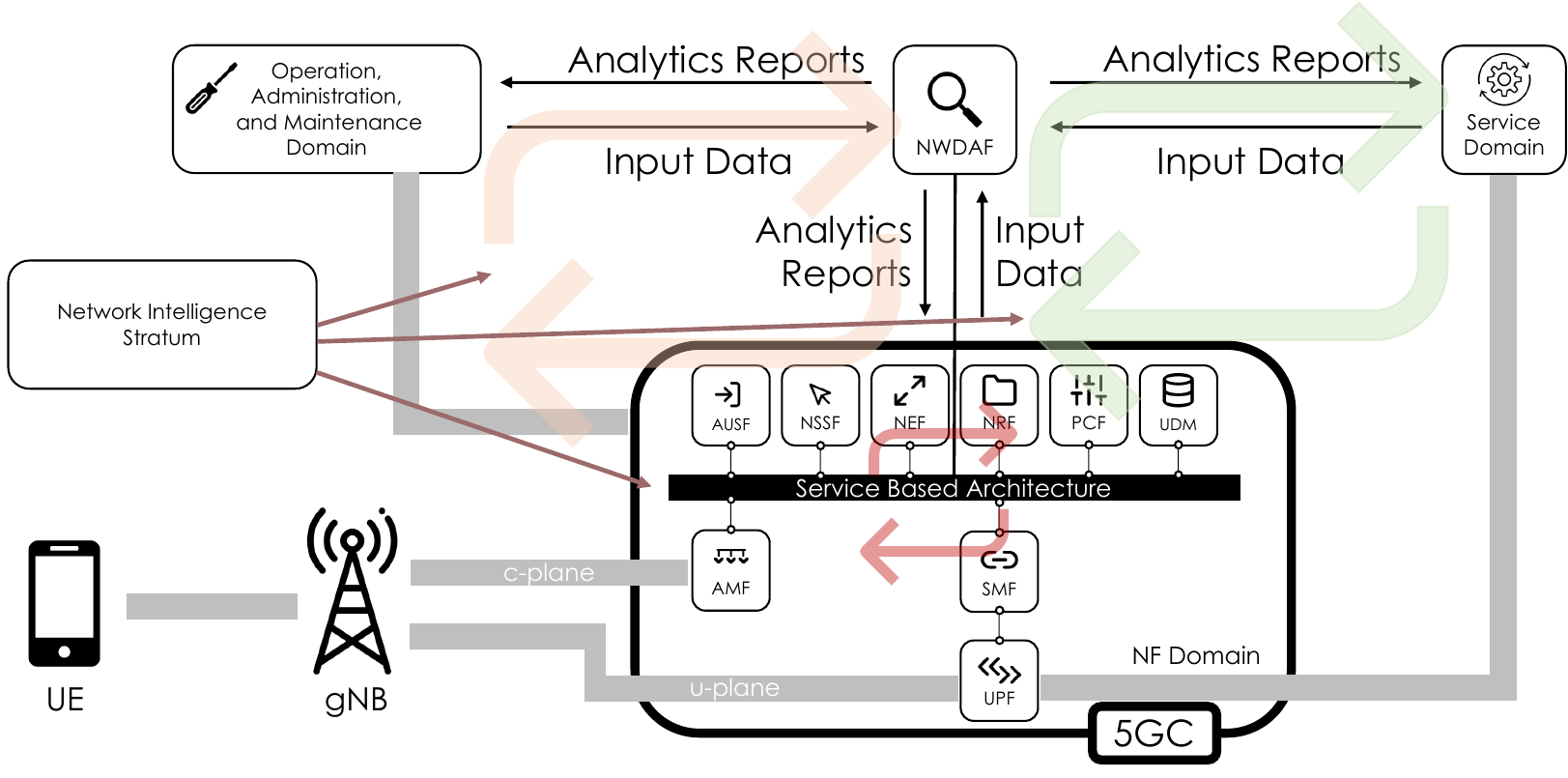}
    \caption{The architectural framework proposed by the 5GPPP Arch WG~\cite{bahare_massod_khorsandi_2023_7313232}.}
    \label{fig:fig13}
\end{figure}

Figure~\ref{fig:fig13} presents the interconnections among various components. The framework is divided into three domains and shows where the \ac{NISt} takes a role. The first domain, referred to as \ac{5GC}, is where the \ac{NWDAF} resides. Within this domain, other \acp{NF} of the core act as the primary producers and consumers of data and analytics. These \acp{NF} utilize the data and analytics to drive network operations in a data-driven manner. Thanks to the \ac{NWDAF}, consumer \acp{NF} no longer need to directly communicate with every potential producer to compute analytics, as they can efficiently leverage the shared information. \ac{NWDAF} is a specific (and very important) \ac{NIF}, that can leverage on a number of \ac{NIF-C} according to the analytics that are served.

The second domain encompasses \ac{OAM} activities, which involve modules such as Element Managers or Network Elements in pre-5G networks. Starting from R15, \ac{OAM} effectively enforces network slicing through the service-based management architecture. The \ac{OAM} domain can also supply the \ac{NWDAF} with data from the \ac{RAN} and 5G \acp{NF}, such as resource consumption. Unlike the pre-5G \ac{3GPP} \ac{RAN} architecture, which lacks an analytics hub like the \ac{NWDAF}, alternative architectures like \ac{O-RAN} feature dedicated analytics modules. The \ac{MDAF} serves as the module responsible for interacting with the \ac{NWDAF} and provides \ac{MDAS}. As discussed, the \ac{MDAF} collaborates with the \ac{NWDAF} and other core \acp{NF} to generate management analytics information, which is subsequently consumed by other \acp{NF} or management procedures like the self-organizing network. From the perspective of the \ac{NISt}, the \ac{MDAF} is an \ac{NIF} that can be further split into several \ac{NIF-C} which (i) interact with the \ac{NWDAF}, effectively closing the loop with the core, and (ii) allows the internal interaction within the management domain.

The third domain encompasses the service domain, facilitated through the \ac{AF}. These functions outside the \ac{3GPP} trust domain play a crucial role in facilitating close interaction between service providers and network operators. This interaction is achieved through enriched service layers, which aid in commoditizing the network and enhancing the interplay between the service and network intelligence. Given the criticality of authorization and security, verifying whether \acp{AF} are appropriately authorized to interact with the \ac{NWDAF} and engage in data exchange with third parties is essential. Authentication can be managed in three different ways. One is basic user-password authentication, where credentials are configured via a configuration file. Support of \ac{TLS} protocol where there is a server-side authentication or mutual \ac{TLS} authentication, where both server-side and client-side authentication is required. In this case, the \ac{AF} can be seen as a specific \ac{NIF-C} (either \textit{Sink} or \textit{Source}, depending on the context). Overall, any \ac{NF} deployed within the \ac{5GC}, the \ac{OAM} system, or any \ac{AF} can contribute input to the \ac{NWDAF} and request analytic reports from it. This establishes a feedback loop where any \ac{NF}, \ac{OAM} component, or \ac{AF} can provide input data to the \ac{NWDAF} and receive analytic reports generated from the collective data obtained by the \ac{NWDAF}. Through these feedback loops, the majority of automated network operations can be executed, as exemplified by the ones already provided by the \ac{NWDAF} in the standard.

\section{Network Intelligence Stratum Procedures}\label{sec:procedures}
In this section, we show how the architectural building blocks that compose the \ac{NISt} will interact with each other. Since the main component of the \ac{NISt} is the \ac{NIO}, we will show how its internal functionalities cooperate to solve the challenges mentioned in Section~\ref{sec:architecture}. Additionally, we will explore how the different components work together to create a cohesive system that can effectively orchestrate intelligence across multiple domains. Through these interactions, the \ac{NIO} will be able to address the challenges that can emerge when \acp{NIS} are deployed across different network domains and operating in multiple timescales.

Notice that all the procedures mentioned below are depicted using a process view. This view answers how the system behaves, addressing concurrency and synchronization aspects. Unified Modeling Language (UML) sequence diagrams were selected as the most appropriate form. Next, we briefly describe how combining some functional blocks can help address the challenges described in the previous section.

\subsection{Inter NIO Procedures} \label{sec:procedures:external}
One of the most essential management and orchestration capabilities is to handle the lifecycle of each of its entities. The \ac{NIO} is not an exception. Regarding networking functionalities, \ac{NFV} \ac{MANO}~\cite{etsi-nfv-mano} is the referent architectural framework to look up to. Lifecycle management is generally responsible for the following operations: creation, instantiation or deployment, management (e.g., model selection and optimization), and termination. However, given the intelligent nature of the \ac{NI}, several factors must be considered while addressing their lifecycle management. In the following subsections, we will discuss in detail how the \ac{NIO} performs lifecycle management of the different \ac{NI}. 

\subsubsection{NI Creation}
When creating a new \ac{NIS}, the \ac{NIO} should verify that all the \acp{NIF} from that \ac{NIS} are available in the catalog. If a \ac{NIF} is unavailable, a new training should be started, e.g., based on user-defined \ac{NIFD}/ \ac{NISD}. This training is represented by triggering a new \ac{MLOps} pipeline. The data ingestion for training this new \ac{NIF} should be coordinated between the \ac{NIO} and the \ac{MLOps} pipeline. Notice that this procedure only contemplates the creation of the \ac{NIF} and not its usage.

\begin{figure}
    \centering
    \includegraphics[width=\columnwidth]{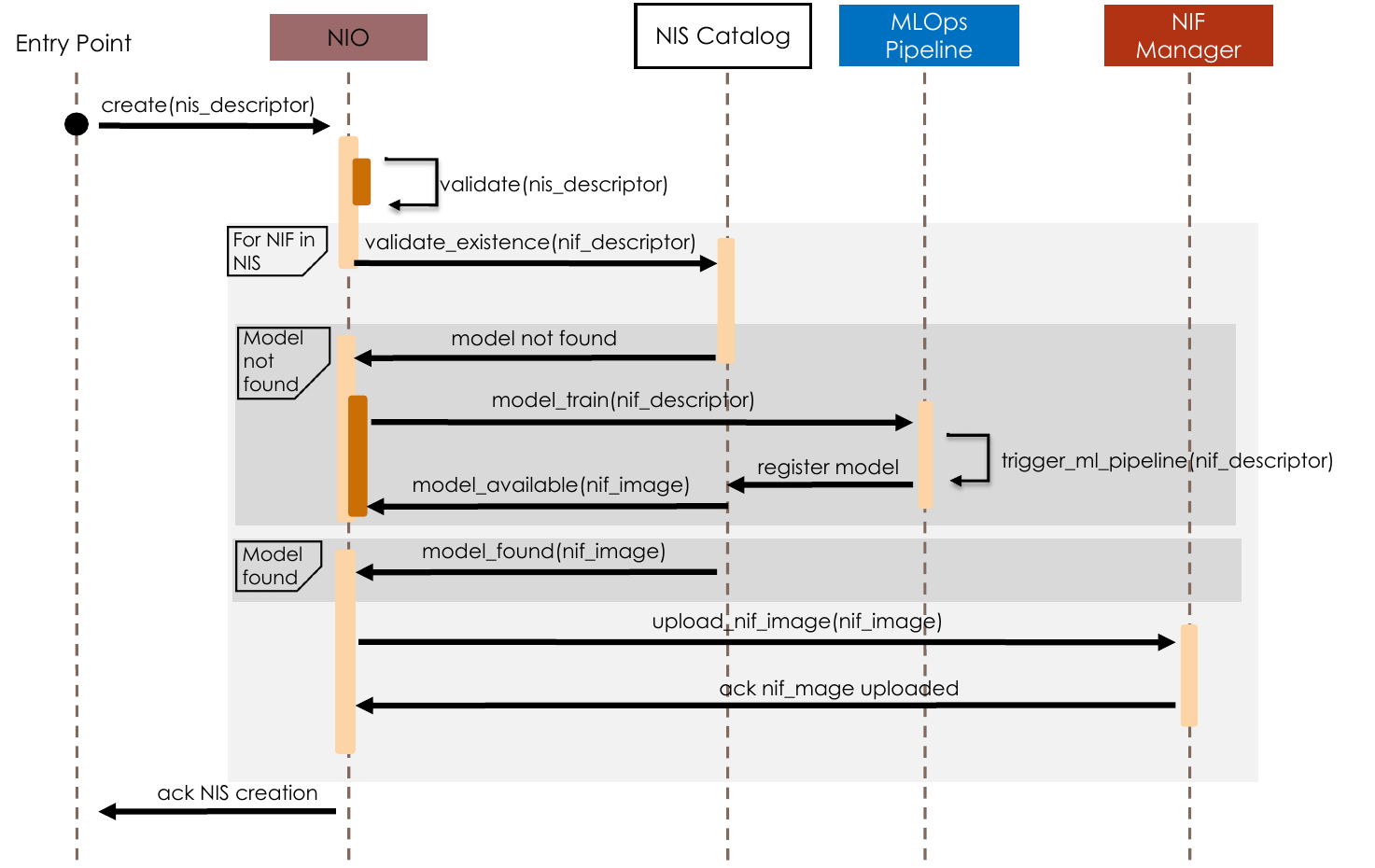}
    \caption{NIS creation process flow.}
    \label{fig:fig14}
\end{figure}

Figure~\ref{fig:fig14} shows the required interactions to create a \ac{NIS}/\ac{NIF}. In the first step, the \ac{NIO} should process a \ac{NIS}/\ac{NIF} creation request through its API. A sender can submit this request, which could be a human, an AI agent, or another process with administration rights to trigger orchestration operations in the \ac{NIO}. The sender identifies that a new \ac{NIS}/\ac{NIF} is needed to perform a given network operation and submits this request to the \ac{NIO}. As input for this process, the \ac{NIO} should receive a \ac{NIFD}/ \ac{NISD} which includes, but is not limited to:

\begin{itemize}
    \item 	Learning mode, if the \ac{ML} model supports online learning or if the training is made offline. 
    
    \item 	Data on which the model is trained (whether the learning is online or offline). This field also specifies the format in which the input data is expected. 
    
    \item 	Learning metrics. This typically includes accuracy, cross-entropy, or a known loss function, e.g., Mean Squared Error (MSE). 
    
    \item 	Model performance upper and lower thresholds. Values on which the training can be concluded (upper threshold). It is assumed that once the upper threshold is met, the \ac{ML} model is ready to be deployed in production. On the contrary, if the lower threshold is met, the \ac{ML} model deployed in production should be updated. The definition of these thresholds may vary depending on the \ac{NI}, but it should reflect the expected performance of the \ac{NI}. 
    
    \item 	Output format. This field specifies the format the \ac{ML} will communicate its output. For instance, a classification problem can produce a vector with the probability of a given sample belonging to a class or the class itself. 
    
    \item 	Last modification time. This field will indicate the age of the \ac{ML} model. Given the constant evolution of network state and data, having an up-to-date \ac{ML} model is crucial for network operation. 
    
    \item	Dependencies required for operation. \ac{ML} models are created using specific libraries (e.g., NumPy, pandas, etc.). The right versions of such libraries must be available when instantiating the \ac{ML} model in production. 
\end{itemize}

As a second step, the	\ac{NIO} then processes the \ac{NIFD}/\ac{NISD}, by checking for the existence of mandatory elements (i.e., network operation, data requirements, output format, and accuracy) and validating the integrity and authenticity of the \ac{NIFD}/ \ac{NISD}. Afterwards, for every \ac{NIF} in the \ac{NIS}, the \ac{NIO} verifies if the \ac{NIF} model exists in the catalog. Two things may happen. If the \ac{NIF} model is not present in the catalog, the \ac{NIO} triggers a training operation from the \ac{ML} pipeline resulting in the execution of a new data ingestion - model training - model testing - model packaging - model registering pipeline. Most of the data needed to execute this pipeline is provided in the \ac{NIFD}. Once the pipeline is completed, a new image from the \ac{NIF} model is registered in the \ac{NIF} Catalog. If the model is present in the catalog, it can be used in inference. For doing this, the \ac{NIO} makes the \ac{NIS}/\ac{NIF} images available to each applicable \ac{NIF-C} Manager. The \ac{NIF}-C Manager acknowledges successful image uploading. Finally, the \ac{NIO} acknowledges the \ac{NIS}/\ac{NIF} creation to the sender. 

\subsubsection{Instantiation or Deployment}
\begin{figure}
    \centering
    \includegraphics[width=\columnwidth]{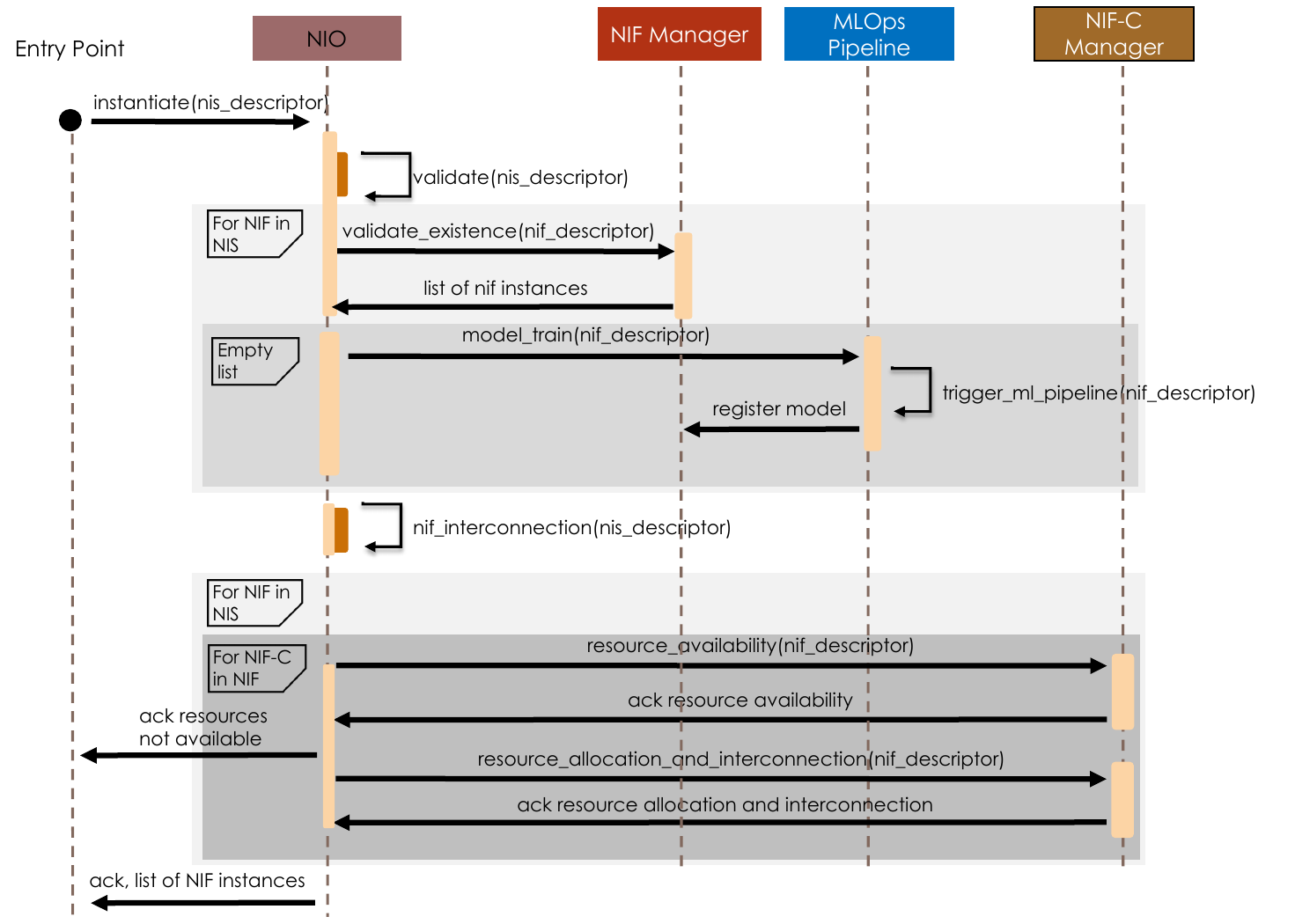}
    \caption{NIS instantiation process flow.}
    \label{fig:fig15}
\end{figure}

Figure~\ref{fig:fig15} shows the interactions required for instantiating or deploying a \ac{NIS}/\ac{NIF}. As in the previous step, the \ac{NIO} receives a request to instantiate a new \ac{NIS}. Then, several variants might be possible.  If none of the \acp{NIF} belonging to the \ac{NIS} is instantiated or deployed, the  \ac{NIS} instantiation will also include the instantiation of all the needed \ac{NIF} instances through the \ac{NIF} Manager. If all the needed \ac{NIF} instances have already been created, the \ac{NIS} instantiation would only deal with the interconnection of the corresponding \ac{NIF} instances. Lastly, a combination of the above is possible where some \ac{NIF} instances might exist, some might need to be created, and instantiated, and some network connectivity between the \acp{NIF} may already exist. 

It is important to notice that if a \ac{NIF} instance is already created, it can be shared between different \acp{NIS}. In this case, the \ac{NIO} should trigger the conflict resolution mechanism because they may be deployed on the same node and/or accessing the same resources. If no conflict is produced, the same \ac{NIF} can instantiate the current \ac{NIS}. If a potential conflict is detected, the \ac{NIO} should proactively address it by deploying specific policies implementing rules or priorities (c.f., Section~\ref{sec:procedures:internal}) to effectively solve the aforementioned conflict.

The main steps for \ac{NIS}/\ac{NIF} instantiation are as follows. First, the \ac{NIO} receives a request to instantiate a new \ac{NIS}/\ac{NIF}. The \ac{NIO} validates the request in terms of the request's validity, including validating that the sender is authorized to issue this request and validation of the parameters passed for technical correctness and policy conformance. For each \ac{NIF} in the \ac{NIS}, the \ac{NIO} checks with the \ac{NIF} Manager if an instance matching the requirements already exists. If such an instance exists, it will be used as part of the \ac{NIS}. If the \ac{NIF} instance does not exist, the \ac{NIO} triggers the \ac{NIF} creation procedure. 

The \ac{NIO} then should perform a feasibility check of the \ac{NIF} interconnection setup. For doing this, the \ac{NIO} requests to the \ac{NIF-C} Manager the availability of resources needed for the \ac{NIF} interconnection and reservation of those resources. The \ac{NIF-C} Manager checks the availability of resources needed for the \ac{NIF} interconnection and reserves them. The \ac{NIF}-C Manager returns the reservation result to \ac{NIO}. If the resources are not available, the \ac{NIS} might not be instantiated, which results in a denial of the \ac{NIS} instantiation. However, if the resources are available, the \ac{NIO} requests the \ac{NIF}-C Manager to allocate and interconnect the \ac{NIF} instances.  The \ac{NIF}-C Manager instantiates the connectivity network needed for the \ac{NIS} and finalizes with a completion acknowledgment. Finally, the \ac{NIO} acknowledges the completion of the \ac{NIS} instantiation.

\subsubsection{Management}
Several operations can be considered as management procedures, such as \ac{NIS}/\ac{NIF} update, optimization, scaling, or migrating. \ac{NI} solutions stored in the \ac{NIS}/\ac{NIF} catalog are inherently trained on hardware and software platforms that may not match the ones available in the new environment where they need to be deployed. In such cases, the \ac{NIS} creation/selection, optimization, and instantiation block will obtain networking and execution context information from its \ac{MANO} block operating in the network and select the proper model to be used in inference within a \ac{NIF}. Suppose a mismatch between trained and targeted hardware/software appears. In that case, the same block should perform the optimization/adaptation (e.g., compression of a neural network, change of inference library from GPU to CPU, etc.) to match the new environment. In case no model is available for the specific execution environment, the \ac{NIS} creation/selection, optimization, and instantiation block will create a new \ac{NIS} and then notify the \ac{NIS} workflow configuration block to trigger a new training phase. Here, we present the \ac{NIS}/\ac{NIF} update with model selection as the most relevant and generic procedure that may involve optimization, re-training, or selection.

\begin{figure}
    \centering
    \includegraphics[width=\columnwidth]{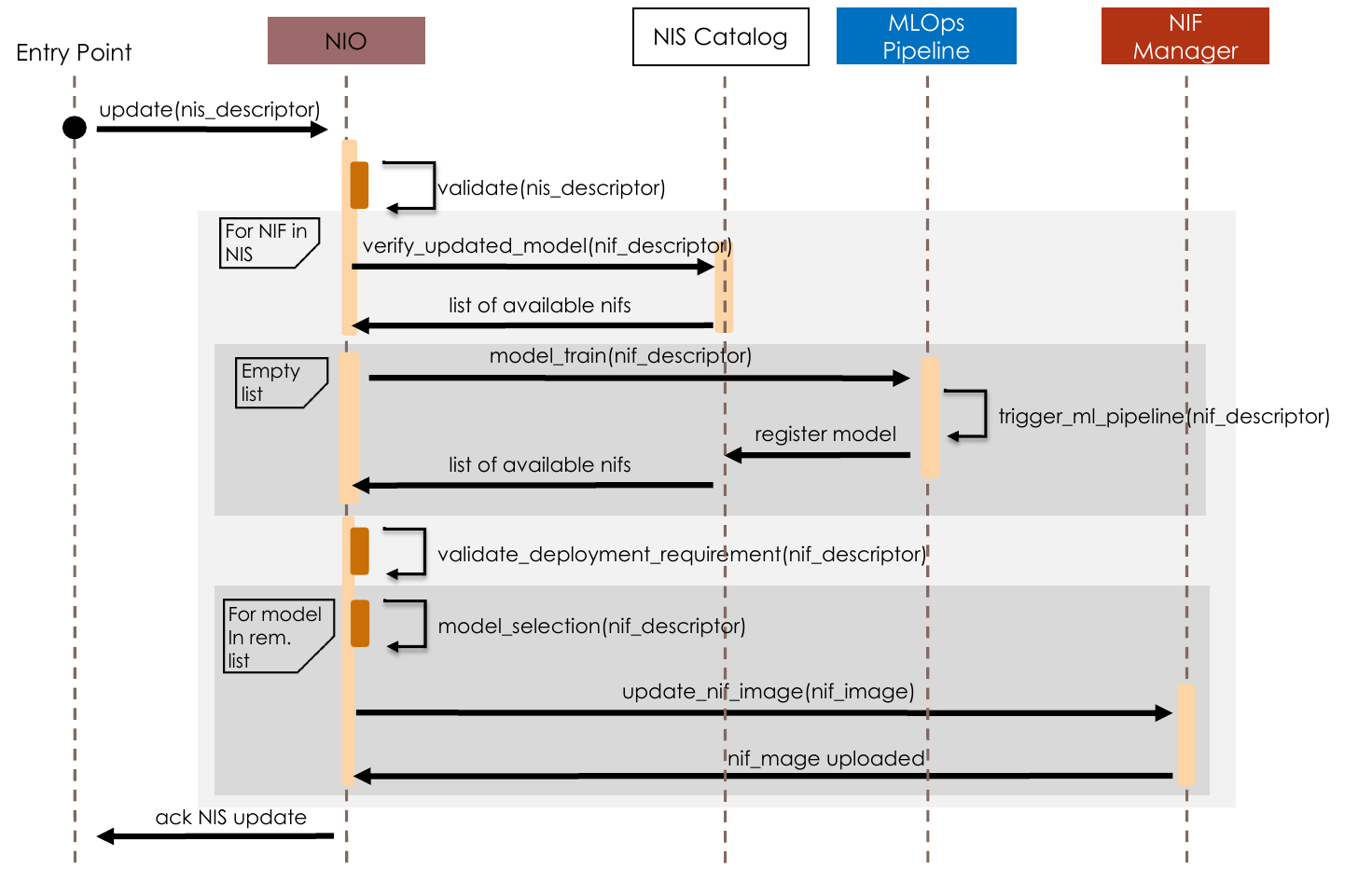}
    \caption{NIS update process flow.}
    \label{fig:fig16}
\end{figure}

Figure~\ref{fig:fig16} shows the main steps for \ac{NIS}/\ac{NIF} updates. This procedure includes updating the parameters of the \ac{NIS}/\ac{NIF}. It is important to notice that the update process has similarities with the \ac{NIS}/\ac{NIF} creation. A request for \ac{NIS}/\ac{NIF} update is submitted from a sender, which could be a human, an \ac{AI} agent, or another process in the architecture, such as a data analytics module that is detecting a mismatch of the statistics of the input data, or a monitoring module detecting that the current model's accuracy is lower than expected. The sender identifies that a new \ac{NIS}/\ac{NIF} needs to be updated and submits its request for an update through the \ac{NIO} \ac{API}. Then, the \ac{NIO} processes the \ac{NIFD}/\ac{NISD} to check the existence of mandatory elements and validate the integrity and authenticity of the descriptor.
   
For every \ac{NIF} in the \ac{NIS} that must be updated, the \ac{NIO} verifies that a new version of the \ac{NIF} model exists in the catalog. Similarly as in the \ac{NIS} creation, if an updated model is needed but is not available in the catalog, the \ac{NIO} triggers a re-training operation from the \ac{MLOps} pipeline, starting a new pipeline. Once the model is re-trained, an image is registered in the \ac{NIF} catalog. Consequently, the \ac{NIFD} is updated with the new version and requirements (i.e., data format, hardware, software dependencies, etc.). On the contrary, if a model (or more than one model) is available, then the \ac{NIO} verifies that the available models satisfy the deployment requirements in terms of data (e.g., input rate and format), computation platform (e.g., CPU, GPU, TPU or FPGA), dependencies (e.g., TensorFlow, PyTorch, etc.), and performance level. This process might return an empty list, meaning that no model satisfies the deployment requirement, and creating a new \ac{NIF} is needed.

If more than one model satisfies the deployment requirements, model selection should be carried out. In this phase, the component will compute an \ac{ML} test score, and depending on arbitration policies, the best-performing model is selected to update the \ac{NIF} image. The \ac{ML} test score can contain learning-related metrics (e.g., loss/reward function) and non-learning-related metrics (e.g., \ac{QoE}, \ac{QoS}, or stability in deployment). The arbitration policies are decision factors that the \ac{NIO} considers primordial for model deployment, for instance, if model precision is preferred over energy consumption. If the model is updated, it should be registered in the catalog and can be used in inference.

Finally, once all the \acp{NIF} that compose the \ac{NIS} are available, the \ac{NIO} makes \ac{NIS}/\ac{NIF} images available to each applicable \ac{NIF}-C Manager. Then, the \ac{NIF-C} Manager acknowledges the successful uploading of the image, and finally, the \ac{NIO} acknowledges the \ac{NIS}/\ac{NIF} update to the sender. Other management operations include optimization, scaling in/out, or migrating. The workflows are similar to those of \ac{NFV}-\ac{MANO}~\cite{etsi-nfv-mano}, requiring an extra step to update the \ac{NIS}/\ac{NIF}, which was shown above.

\subsubsection{Termination}
The request for terminating a \ac{NIS}/\ac{NIF} is received by the \ac{NIO}. This request might come from a human, an \ac{AI} agent, or another process in the architecture. When terminating a \ac{NIS}/\ac{NIF} instance, several variants might be possible. In case all \ac{NIF} instances contributing to the \ac{NIS} need to be terminated, a termination procedure is started for all the \ac{NIF}, including the removal of the interconnectivity between these \ac{NIF}. In case some \ac{NIF} instances are contributing to other \ac{NIS} instances, only those \acp{NIF} that do not contribute to other \ac{NIS} instances must be terminated. The interconnectivity between them must be removed, leaving the other \ac{NIF} instances in place and the interconnectivity between them intact. 

For terminating a \ac{NIS}/\ac{NIF} instance, the \ac{NIO} receives a request to terminate a \ac{NIS}/\ac{NIF} instance using the \ac{NIS}/\ac{NIF} Lifecycle Management interface. As in previous procedures, the \ac{NIO} validates the request. It verifies the validity of the request (including the sender's authorization) and verifies that the \ac{NIS}/\ac{NIF} instance exists. The \ac{NIO} then proceeds to request the \ac{NIF} Manager to terminate any \ac{NIF} instances that were instantiated along with the \ac{NIS} instantiation, provided they are not used by another \ac{NIS}. At the same time, the \ac{NIF} Manager requests the deletion (release) of resources for this \ac{NIF} instance to the \ac{NIF}-C Manager. For all the \ac{NIF-C}, the \ac{NIF-C} Manager deletes (releases) the resources and then acknowledges the completion of resource deletion back to \ac{NIF} Manager. This completes the deletion of the \ac{NIF}. Once the \ac{NIS} is terminated, the \ac{NIF} Manager sends a confirmation to the \ac{NIO} that the \acp{NIF} are terminated.

\subsubsection{Other Operations}
Operations such as deleting, querying, enabling, or disabling a \ac{NIS}/\ac{NIF} are also considered within the architecture defined by the \ac{NISt}. However, such operations are not different than those proposed in \ac{NFV}-\ac{MANO} as they do not involve or require interactions with any NI-related block and the \ac{MANO} block can perform it. The implementation of such procedures is shown in~\cite{etsi-nfv-mano}. 

\subsection{Intra NIO Procedures} \label{sec:procedures:internal}
As introduced above, a \ac{NIS} is usually composed of different \acp{NIF} and hence, some of the \ac{NIS} management functionalities take place only within the \ac{NIO} itself. These intra \ac{NIO} functionalities address the challenges that may emerge when \acp{NIS} are deployed across different network domains and operating in multiple timescales, including conflict resolution and knowledge sharing among \ac{NIS}.

\subsubsection{Conflict Resolution}\label{sec:procedures:internal:conflict_resolution}
We introduced two specific conflict cases in Section~\ref{sec:funct-NI-stratum}: (i) when conflicts emerge when monitoring data, e.g., algorithms may need data from the same source but with different granularity, and (ii) when conflicts in the policy enforcement of different \ac{NI} algorithms may act on the same network functions but configuring different values for the target parameters. In such situations, the policy interpreter and configuration block will gather information about the policy guiding the different \acp{NIS} and pass their interpretation to the conflict detection and resolution module. In both cases, a conflict will be detected, and the \ac{NIO} will identify and apply the conflict resolution rules associated with (i) multi-timescale coordination and (ii) parameter constraints and execution priority. After applying the rules, the outcome should provide a plan that will trigger a configuration modification of the \ac{NIS} policies. In the case of \ac{NIS} empowered by black-box \ac{ML} algorithms, the Model Explainability block will interpret policies associated with such algorithms.
 
\begin{figure}
    \centering
    \includegraphics[width=\columnwidth]{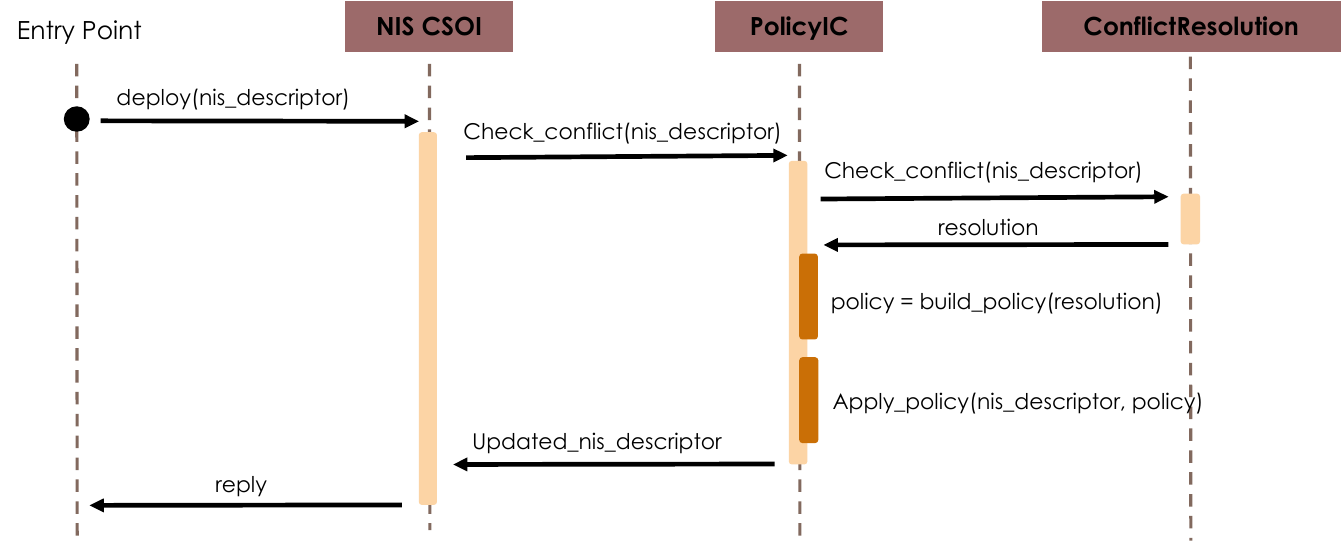}
    \caption{Conflict Resolution process flow.}
    \label{fig:fig17}
\end{figure}

Figure~\ref{fig:fig17} shows the main steps for the case of \ac{NIS} Conflict Resolution. This procedure includes checking the parameters of the \ac{NIS} against the \ac{PolicyIC} to arbitrate the deployment of the \ac{NIS} (e.g., if the \ac{NIS} has different monitoring granularity in a Source shared with other \ac{NIS}, or requires controlling a \ac{NF} that another \ac{NIS} is already controlling with a different \ac{AI} algorithm). When the \ac{NIO} receives a request for the instantiation or updating a \ac{NIS}, it will be the \ac{NIS} \ac{CSOI} component that will internally validate the \ac{NIS}, indicating if there is a conflict and updating and resolving the \ac{NIS} in case any conflict exists. 

The validation command executed in the \ac{NIO} when creating, instantiating, and updating a \ac{NIS} includes the following steps internally. First, the \ac{NIS} \ac{CSOI} validates the \ac{NISD}.	If the \ac{NIS} request is correct and sound, the \ac{NIS} \ac{CSOI} verifies through the \ac{PolicyIC} if there is any conflict by gathering information about the policy guiding the different \ac{NIS}s and passing their interpretation to the Conflict Resolution component. Then, the Conflict Resolution component checks if the \ac{NIS}s to be deployed has any conflict with the existing \ac{NIS}s. The Conflict Resolution component globally solves trade-offs that may emerge from conflicting objectives in the control and user planes, e.g., establishing policies (at small timescales) versus enforcing such policies (at large timescales). For the case of conflicting \acp{NIS}, the Conflict Resolution component compares policies among different \ac{NIS} to detect conflicts that may appear with the new/ updated \ac{NIS}. It performs conflict resolution based on comparison and resolution rules, providing a \ac{NIS} configuration. This configuration will result from a trade-off or priority mechanism that the Conflict Resolution component will execute to harmonize the \acp{NIS}' coexistence. The resolution will contain the last valid configuration if no feasible solution exists.	Once the \ac{PolicyIC} receives the resolution, the new policy is built and applied to the specific \ac{NIS}. Then, the \ac{PolicyIC} returns the \ac{NISD} to the \ac{NIS} \ac{CSOI}. Consequently, the \ac{NIS} \ac{CSOI} further proceeds with the required \ac{NIS} operation (i.e., creation, deployment, update, etc.). Eventually, the \ac{NIO} acknowledges the \ac{NIS} deployment to the sender.
 
\subsubsection{Knowledge Sharing}
 
\begin{figure}
    \centering
    \includegraphics[width=\columnwidth]{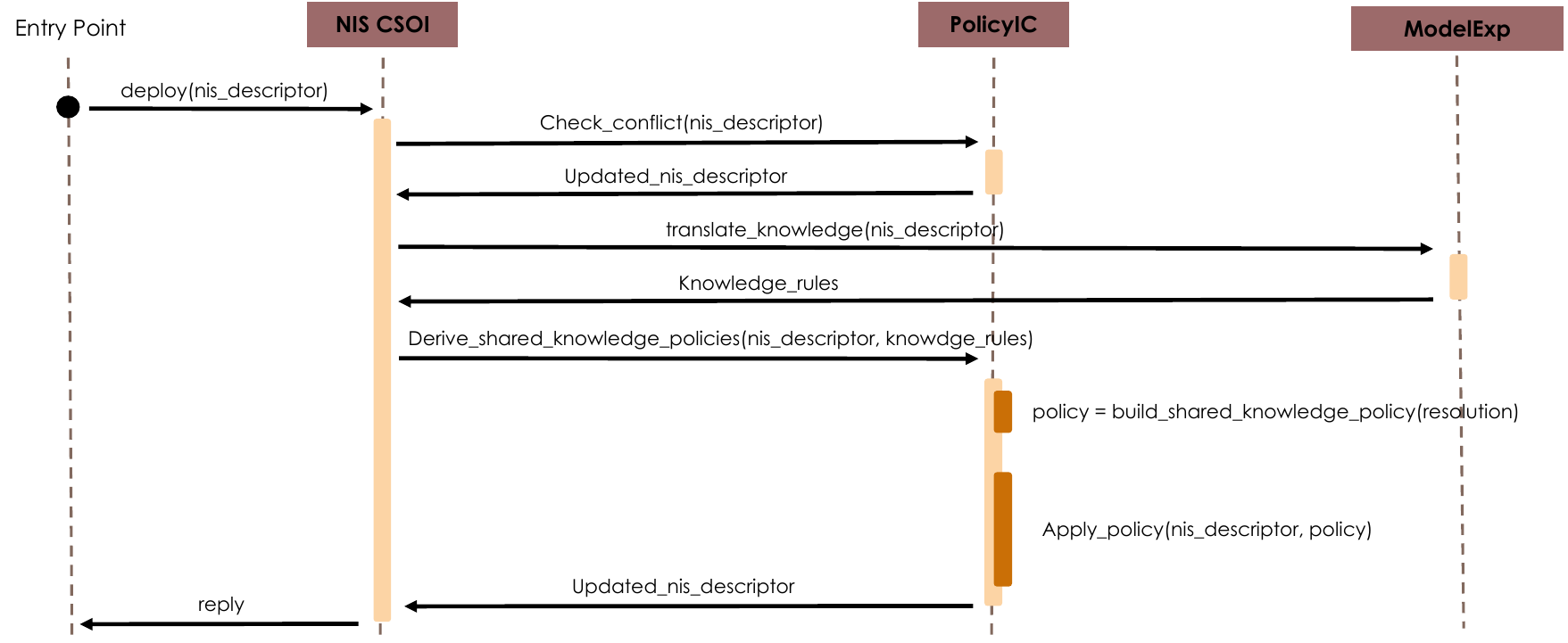}
    \caption{Knowledge Sharing process flow.}
    \label{fig:fig18}
\end{figure}

\acp{NIS} deployed in the same or across different domains use their knowledge to derive their execution plans. The knowledge management block will allow the \ac{NIO} to understand the knowledge of each \acp{NIS}, via the interaction with the Model Explainability block and derive new policies that represent the shared knowledge among \acp{NIS}, by interacting with the \ac{PolicyIC} block.

Figure~\ref{fig:fig18} shows the main steps for the case of \ac{NIS} Knowledge Sharing. This procedure includes checking the parameters of the \ac{NIS} against the \ac{PolicyIC} initially (and consequently also with the Conflict Resolution component internally). However, for the cases in which a \ac{NIS} requires the use of knowledge coming from an external domain, the \ac{NIS} \ac{CSOI} will first translate such knowledge in the Model Explainability block before building and applying the shared knowledge policies:

When a new \ac{NIS} instantiation is requested, the \ac{NIO} will process this request similarly as described in the previous sections, i.e., validation, conflict resolution, and policy update.  With the updated \ac{NISD}, the \ac{NIS} \ac{CSOI} requests the translation of the external domain knowledge to the Model Explainability block. As a result, the \ac{NIS} \ac{CSOI} receives additional Knowledge rules. Then, the \ac{NIS} \ac{CSOI} sends the \ac{NISD} again to the \ac{PolicyIC}, but this time together with Knowledge rules to build and apply the shared knowledge policies. If shared knowledge policies must be built, this is done by the \ac{PolicyIC} module, taking in account possible existing conflicts. When shared knowledge policies must be applied, this is also done by the \ac{PolicyIC} block by returning the \ac{NISD} to the \ac{NIS} \ac{CSOI}. Finally, the \ac{NIO} acknowledges the \ac{NIS} deployment to the sender.

\begin{figure}
    \centering
    \includegraphics[width=\columnwidth]{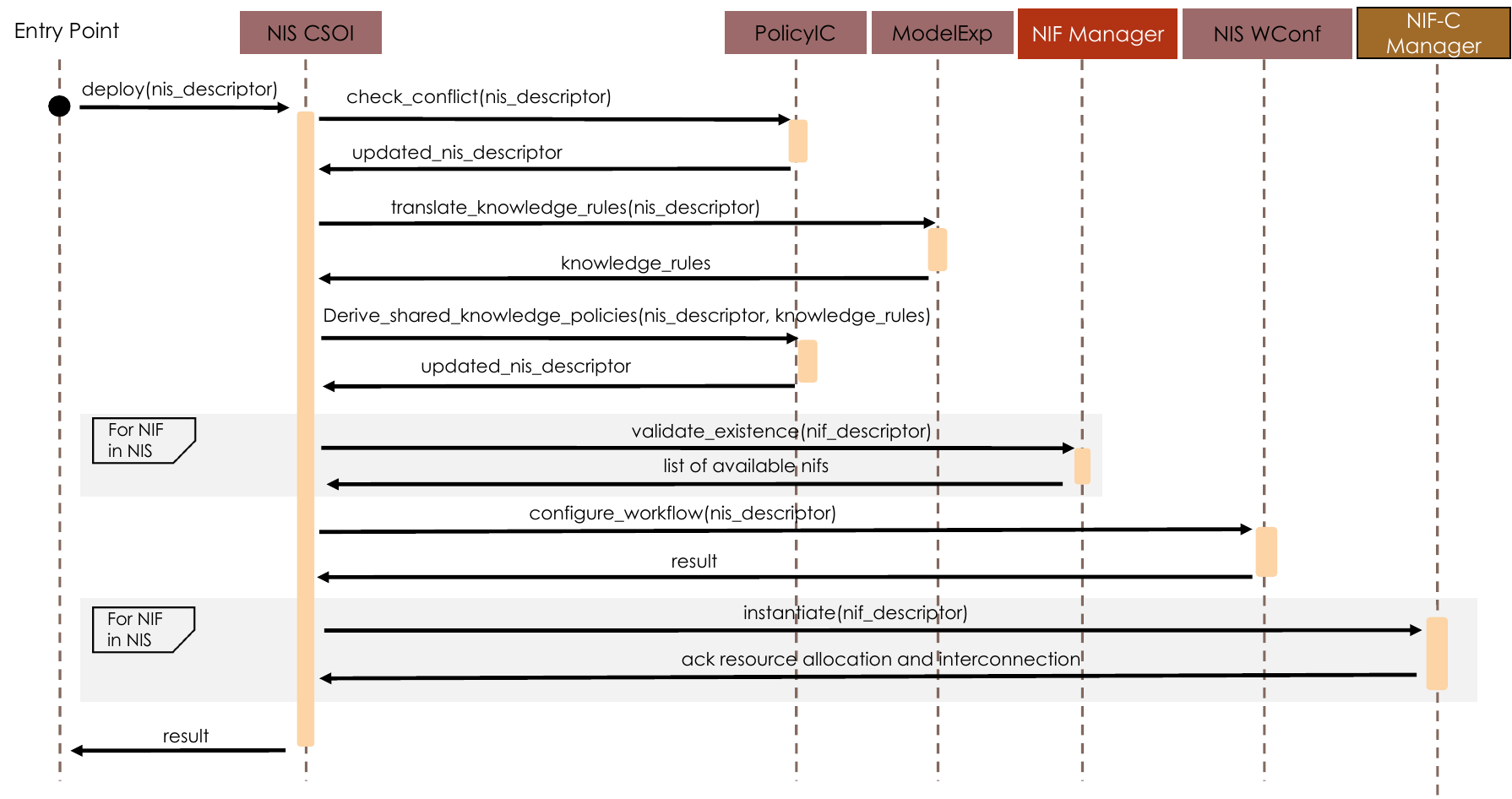}
    \caption{Intra NIO instantiation and deployment process flow.}
    \label{fig:fig19}
\end{figure}

\begin{table*}
\centering
\caption{Summary of the procedures proposed to address the challenges described in Section~\ref{sec:architecture} and the functionalities of the NIO that can be used to achieve it.}
\label{tab:summary-procedures}
\resizebox{\textwidth}{!}{%
\begin{tabular}{|c|c|c|}
\hline
\textbf{Procedure}                     & \textbf{Procedure type} & \textbf{Functional blocks}                                                                                                                                               \\ \hline
Creation                               & Inter NIO Procedures    & NIO, NIS Catalog, ML Pipelines, NIF Manager                                                                                                                              \\ \hline
Instantiation or Deployment            & Inter NIO Procedures    & NIO,  ML Pipelines, NIF Manager, NIF-C Manager                                                                                                                           \\ \hline
Management                             & Inter NIO Procedures    & NIO, NIS Catalog, ML   Pipelines, NIF Manager, MANO                                                                                                                      \\ \hline
Termination                            & Inter NIO Procedures    & NIO,   NIF Manager, NIF-C Manager                                                                                                                                        \\ \hline
Other operations                       & Inter NIO Procedure     & As in NFV-MANO                                                                                                                                                           \\ \hline
Conflict Resolution                    & Intra NIO Procedures    & \begin{tabular}[c]{@{}c@{}}Policy Interpreter and Configuration, \\ Creation Selection Optimization and Instantiation, \\ Conflict Detection and Resolution\end{tabular} \\ \hline
Knowledge Sharing                      & Intra NIO Procedures    & \begin{tabular}[c]{@{}c@{}}Policy Interpreter and Configuration, \\ Model Explainability, \\ Knowledge management\end{tabular}                                           \\ \hline
Intra NIO Instantiation and deployment & Intra NIO Procedures    & \begin{tabular}[c]{@{}c@{}}Policy Interpreter and Configuration, \\ Model Explainability, ML Pipelines, \\ NIF Manager,   \\ NIF-C Manager\end{tabular}                  \\ \hline
\end{tabular}%
}
\end{table*}

\subsubsection{Intra NIO Instantiation and deployment}
The previously described Conflict Resolution and Knowledge Sharing mechanisms are inherent to the \ac{NIS} \ac{CSOI} and Lifecycle Management components interacting with external components such as the \ac{NIS} Catalog, the \ac{NIF} Manager, or the \ac{NIF-C} Manager. To illustrate how the external processes would occur inside the \ac{NIO}, Figure~\ref{fig:fig19} details the deployment interactions between the \ac{NIS} \ac{CSOI} with both internal and external components. The procedure includes the steps to validate the \ac{NISD} and identify and solve possible conflicts before deployment. Also, domain-specific policies are built and applied, followed by training new models in case there are no instances of them already in the catalog. Finally, the \ac{NIS} Workflow Configuration block combines them to build the \ac{NIS} and start the instantiation and deployment. 

The detailed sequence of steps is described as follows. First, a request for deploying a \ac{NIS} is submitted from the sender (it could also be a \ac{NIS} update). The sender identifies that a new \ac{NIS} needs to be deployed and submits its request to the \ac{NIO} through the \ac{NIO} \ac{API}. Then, the \ac{NIS} \ac{CSOI} component receives the request. The \ac{NIS} \ac{CSOI} processes the \ac{NISD}, including, but not limited to:
\begin{itemize}
    \item 	Checking for mandatory elements (i.e., network operation, data requirements, output format, accuracy). 
    \item Validating the integrity and authenticity of the descriptor.
\end{itemize}

If the \ac{NIS} request is correct and sound, the \ac{NIS} \ac{CSOI} will proceed with the validation of the \ac{NIS}.  Please note the validation command executed in the \ac{NIO} for the creation, instantiation, and update processes described in Section~\ref{sec:procedures:external} also includes the procedures of Conflict Resolution and Knowledge Sharing. Then, the \ac{NIS} \ac{CSOI} verifies against the \ac{PolicyIC} if there is still any conflict. As described above, the \ac{PolicyIC} internally requests the Conflict Resolution component to check further if the \ac{NIS}  has any conflict with existing \ac{NIS}. Once the \ac{PolicyIC} receives the resolution, the new \ac{NIS} domain-specific policies are built and applied, and an updated \ac{NISD} is returned to the \ac{NIS} \ac{CSOI}. 

With the updated \ac{NISD}, the \ac{NIS} \ac{CSOI} requests the translation of the external domain knowledge to the Model Explainability block. As a result, the \ac{NIS} \ac{CSOI} receives the additional Knowledge rules. Later on, the \ac{NIS} \ac{CSOI} sends the \ac{NIS} descriptor again to the \ac{PolicyIC}, but this time together with Knowledge rules. Hence, shared knowledge policies are built and applied as previously described in the above sections. The \ac{NIS} \ac{CSOI} now iterates for every \ac{NIF} in the \ac{NIS} and checks if an instance of the given \ac{NIF} already exists in the \ac{NIF} Manager. If no instance exists, a new model will be trained for that \ac{NIF} in the \ac{MLOps} Pipeline.

Next, the \ac{CSOI} proceeds with the interconnection of all \acp{NIF} in the \ac{NIS}. This mechanism involves requesting the \ac{NIS} Workflow Configuration block to virtually link the \acp{NIF} and define their interactions. Finally, the \ac{NIS} \ac{CSOI} starts the instantiation of every \ac{NIF} in the \ac{NIS} in the \ac{NIF-C} Manager. As previously described, this involves checking the resource availability for that \ac{NIF} in the infrastructure by the \ac{NIF-C} Manager. If resources are available, allocate the resources for that \ac{NIF}, and interconnect with the other \acp{NIF} instances through the \ac{NIF-C} Manager. If resources are not available, notify the sender accordingly. Eventually, the \ac{NIO} acknowledges the \ac{NIS} deployment to the sender. 

Table~\ref{tab:summary-procedures} summarizes the functionalities proposed in Section~\ref{sec:architecture} for the \ac{NIO} and which procedures use them. Notice that the procedures described in the previous sections are also based on the related challenges in Section~\ref{sec:architecture}. However, further procedures can be defined based on other use cases, e.g., orchestration of \ac{NI} in federated domains or intelligent orchestration of \acp{NIS}, where the decisions of the \ac{NIO} are empowered by \ac{AI}-based decision-making models). We expect that these procedures can be further extended to more complex cases or used as a reference to define new ones. 

\section{Network Intelligence Stratum Implementation}\label{sec:results}
This section shows a reference implementation of the proposed \ac{NISt}. The first subsection shows how two \acp{NIF} can be described and orchestrated to form a \ac{NIS}. Then, assuming that both \acp{NIF} are acting upon the same network element (i.e., reallocating and scaling the same network service), we show how selected functionalities of the \ac{NISt} can be implemented.  

\begin{figure}[t!]
    \centering
    \includegraphics[width=\columnwidth]{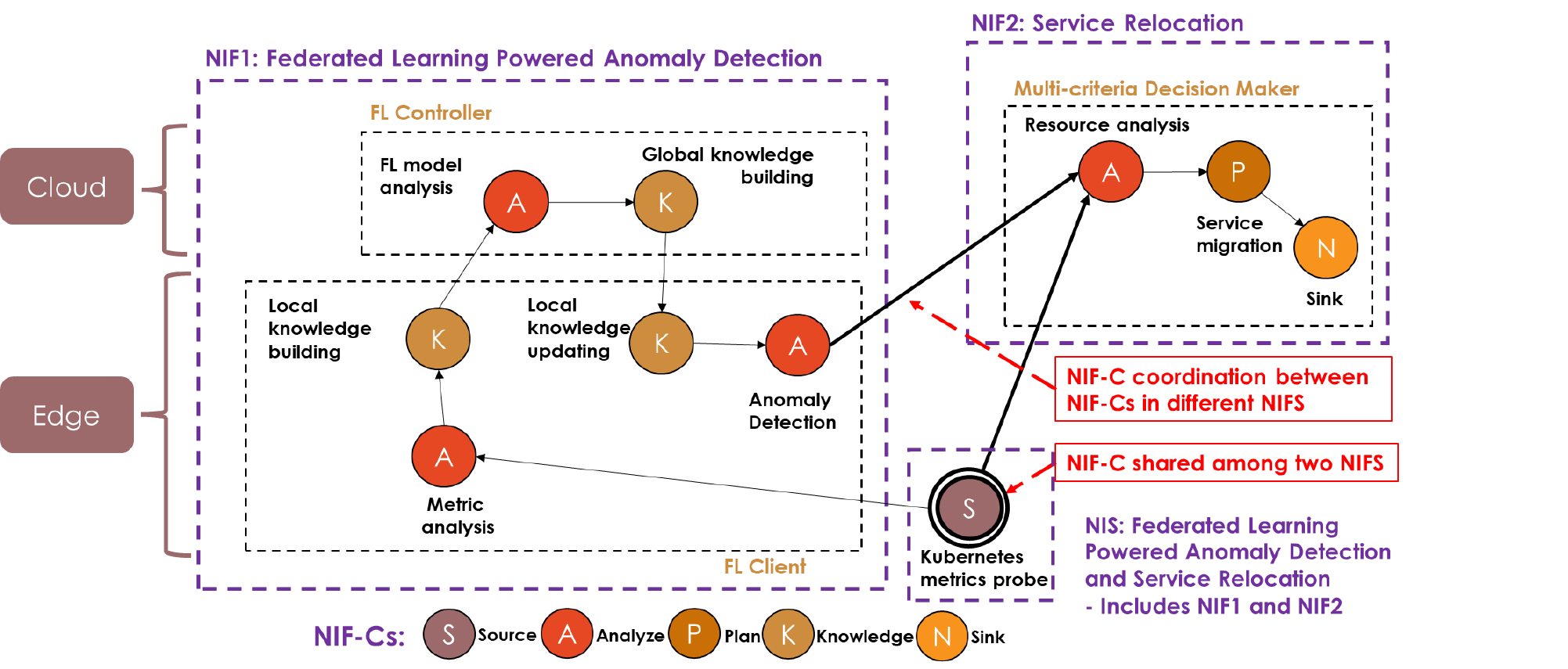}
    \caption{A federated learning-powered anomaly detection and service relocation NIS.}
    \label{fig:fig20}
\end{figure}

\begin{figure*}[t!]
    \centering
    \includegraphics[width=.8\textwidth]{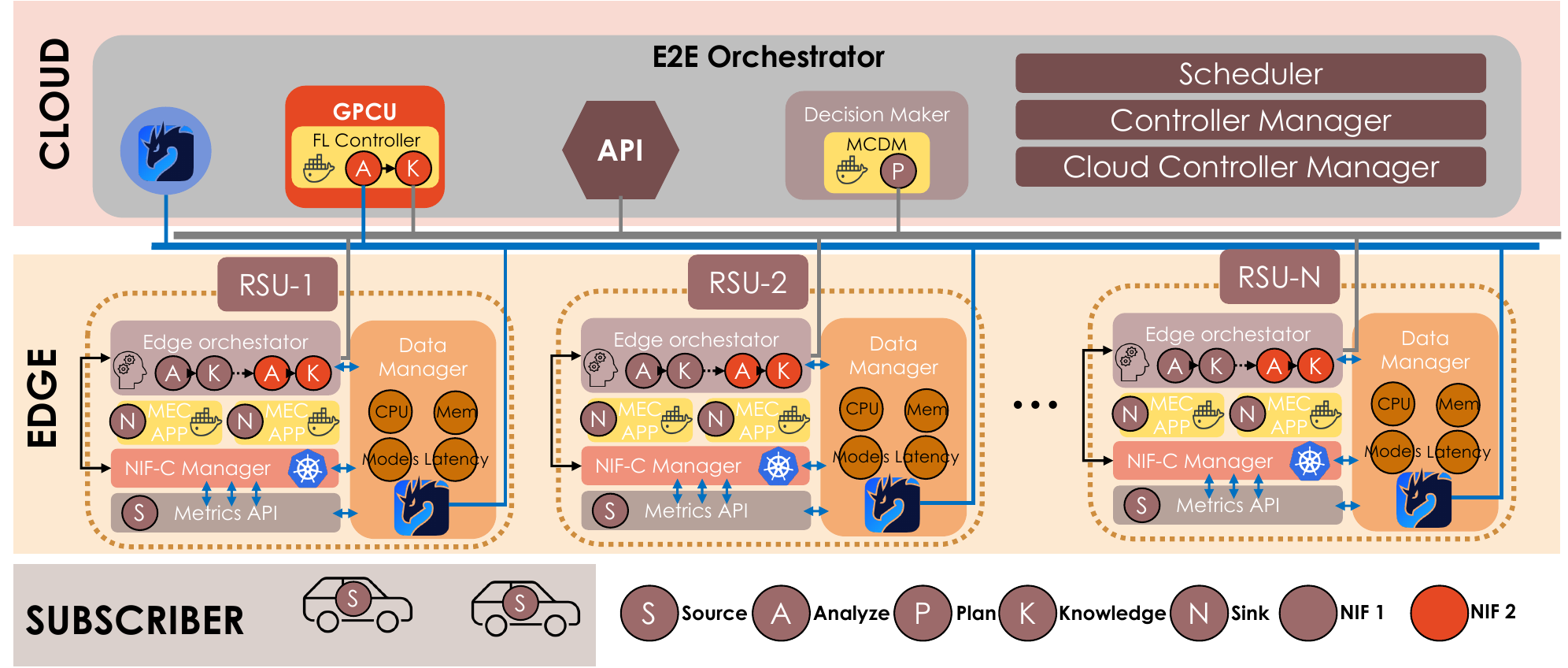}
    \caption{Cloud-to-edge deployment of the proposed NIS and its different components.}
    \label{fig:fig21}
\end{figure*}

\subsection{Combining NIFs to build a NIS} \label{sec:results:orchestrate_implementation}
One of the key features of the \ac{NISt} is to allow the creation, management, and deployment of \acp{NIS}. This section showcases a key functionality of such Stratum: combining two \acp{NIF} to create a \ac{NIS}. The first \ac{NIF} utilizes a federated learning-powered anomaly detection algorithm, enabling anomaly detection at the edge (NIF1 in Figure~\ref{fig:fig20}). This means that the \ac{AI} model for detecting anomalies is trained locally on individual devices, while a central controller is responsible for retaining a global model and, therefore, enhancing the local \ac{AI} model. This way, data privacy is preserved while the approach still benefits from a collaborative learning process. The second \ac{NIF} employs a service relocation algorithm that collects real-time monitoring information from the edge~\cite{slamnik2023ml}, such as CPU load, memory load, used storage, and end-to-end latency measured at the client side. Leveraging a multi-criteria decision-making function, this algorithm dynamically relocates services based on real-time data (NIF2 in Figure~\ref{fig:fig20}), ensuring optimal resource utilization and service performance. It is important to note that by service relocation, we mean relocation of stateless services, a procedure that deploys the correct service instance on the selected edge and terminates all previous service instances running on the other edges to save the edge resources.

Following the proposed framework based on the \ac{N-MAPE-K} to define \ac{NIS}, we effectively combined the federated learning-powered anomaly detection with service relocation capabilities to achieve the \ac{NIS}. Figure~\ref{fig:fig20} shows the \ac{N-MAPE-K} based diagram of the resulting \ac{NIS}. By integrating the federated learning approach, the \ac{NIS} ensures that anomalies are detected securely and efficiently across the network's edge devices. The \ac{NIS} also leverages the monitoring information collected from the edge to make informed decisions about service relocation, maximizing the network's overall performance and responsiveness.

\begin{figure}[t!]
    \centering
    \includegraphics[width=\columnwidth]{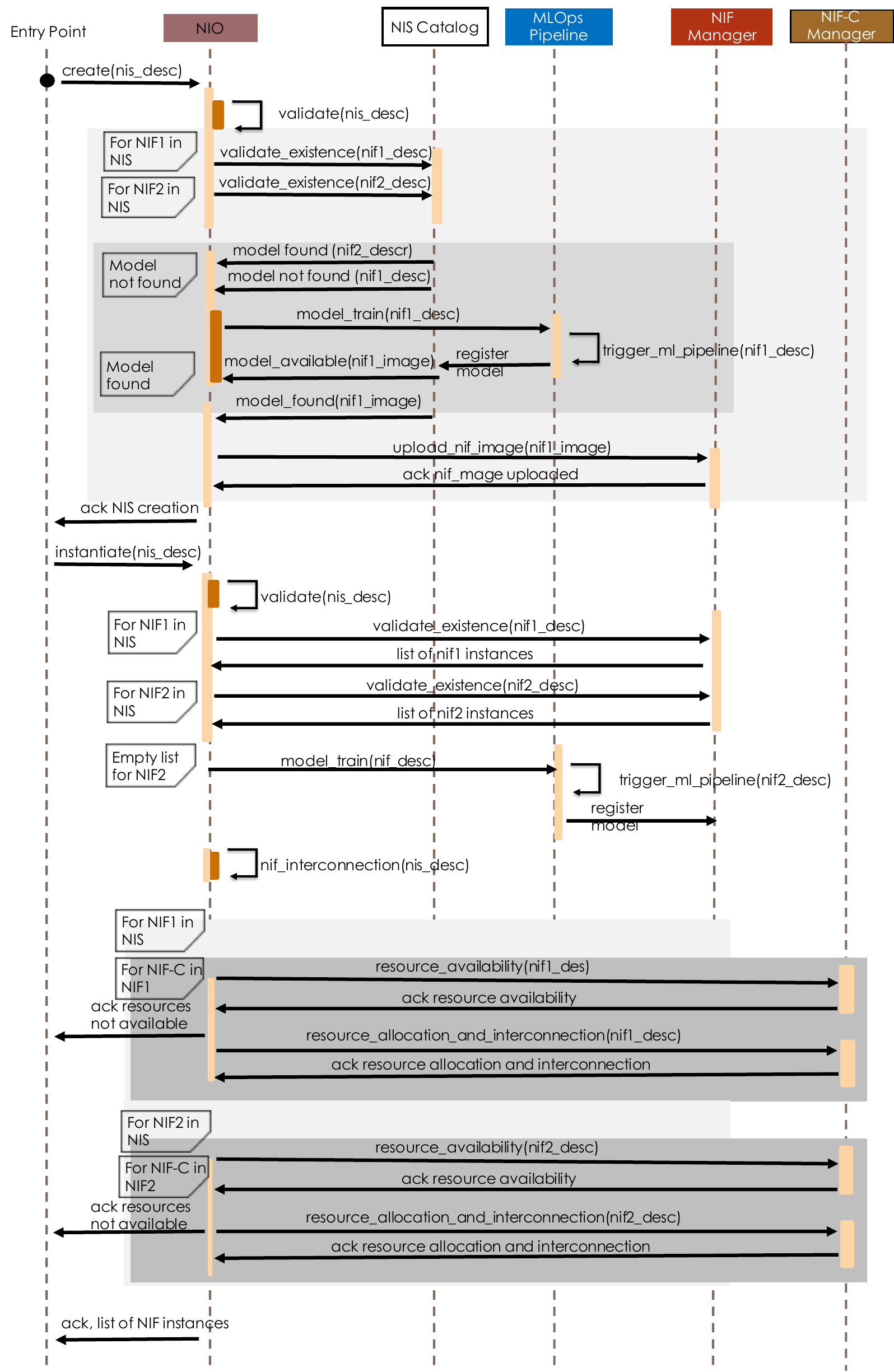}
    \caption{NI Procedures for the example NIS.}
    \label{fig:nis_nif1_nif2_combo}
\end{figure}

Before implementation, the \ac{NIS} needs to be created, which is a procedure that includes validating the existence of constituting \acp{NIF}, and if one of them is not found, the proper models are created and trained, resulting in \ac{NIF} images that are necessary for \ac{NIS} implementation (Figure~\ref{fig:nis_nif1_nif2_combo}). The communication among the \ac{NIS} is implemented using the Eclipse Zenoh data communication framework~\cite{zenoh}, which provides a reliable and scalable solution for exchanging data between devices and components within the network. Additionally, the implementation utilizes Kubernetes~\cite{kubernetes} to realize the \ac{NIF} Manager and the \ac{NIF-C} Manager, as shown in Figure~\ref{fig:fig21}. Kubernetes helps manage the deployment, scaling, and orchestration of the \acp{NIF-C}, ensuring smooth operation and efficient service relocation. 

The Smart Highway testbed~\cite{smarthighway} located on top of the E313 highway in Belgium served as the edge environment for testing and validating the effectiveness of the proposed \ac{NIS}, providing a real-world scenario to assess its performance and potential benefits for vehicular services. Figure~\ref{fig:fig21} shows a high-level representation of the deployment and how the components were deployed at the cloud  (centralized server) and edge (roadside units at the Smart Highway). As shown in Figure~\ref{fig:fig21}, Zenoh-based data managers are used for collecting metrics related to edge performance (CPU, memory, storage) and service performance (end-to-end latency). The edge performance metrics are used in the anomaly detection process (NIF1), while both edge metrics and service metrics are used in the service relocation process (NIF2). In the latter case, the service performance is measured as end-to-end latency collected from the client. The client is implemented on the test vehicle, which measures latency in communication with services deployed at the network edges in real-time. Once the final decision on the service relocation is made, i.e., when \ac{NIS} decides to which network service the test vehicle should connect, the subscriber running in the vehicle receives all necessary information for connecting to the service. 

The outcome of the deployment of this \ac{NIS} is an edge-aware service relocation, which makes efficient decisions on when and where the service should be relocated to ensure minimum end-to-end latency at every moment, taking into account not only the edge performance but also its stability in terms of possible anomalies. Such intelligent services pave the way towards more reliable and high-performing deployments of vertical services in future 6G ecosystems. Nevertheless, the two \acp{NIF} presented in this section could produce conflicting decisions, which might severely affect the overall \ac{NIS} performance. Thus, it is essential to ensure proper conflict resolution, as explained in Section~\ref{sec:results:conflict_resolution}.

\begin{figure*}
    \centering
    \includegraphics[width=0.9\textwidth]{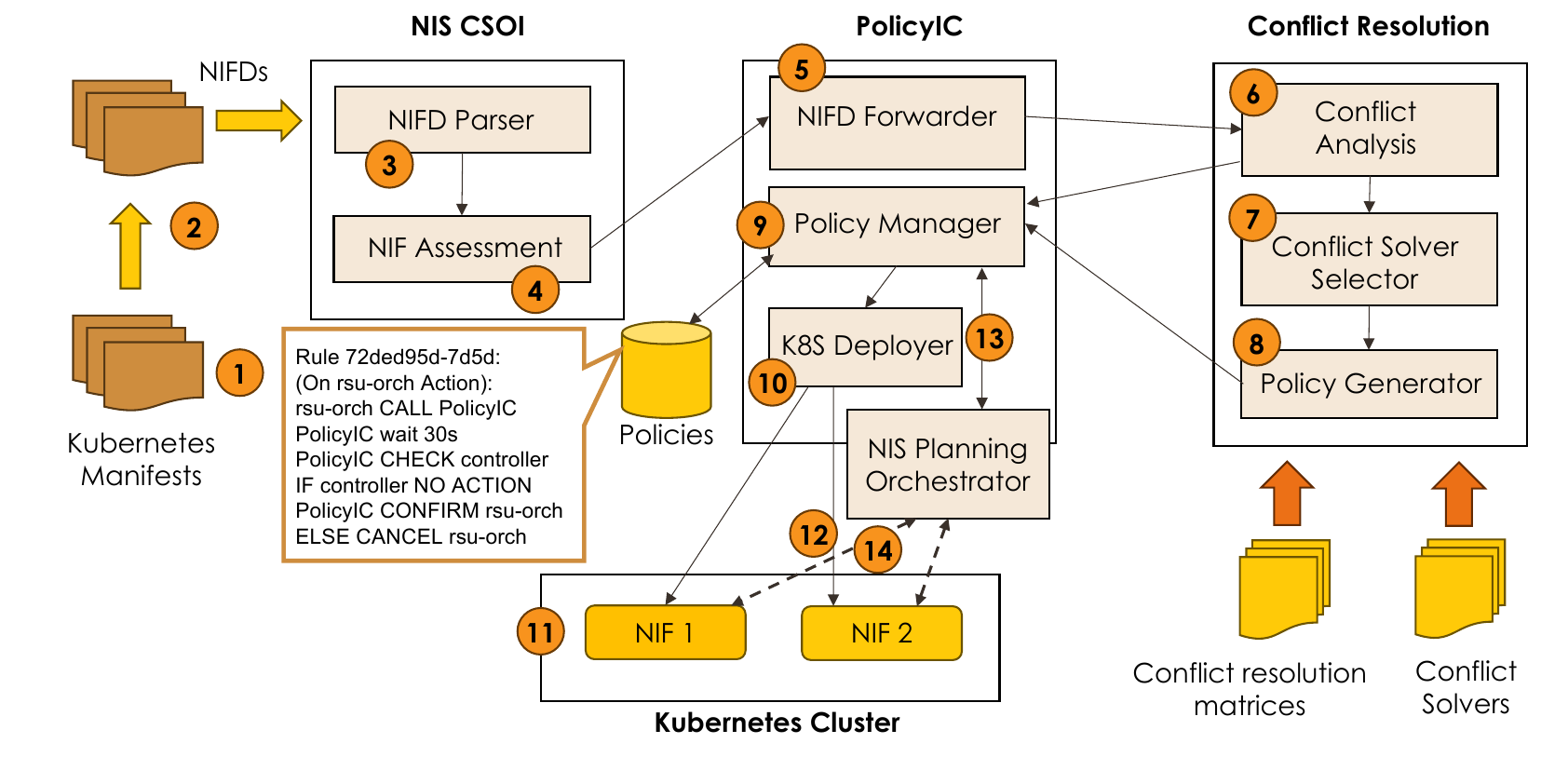}
    \caption{Conflict resolution process.}
    \label{fig:conflict_resolution_implementation}
\end{figure*}

\subsection{Managing a conflicting NIS}\label{sec:results:conflict_resolution}

The conflict resolution procedure is one of the \ac{NISt} procedures with paramount importance for the stability and smooth operation of future 6G networks. The architectural details of the procedure are described in detail in Section~\ref{sec:procedures:internal:conflict_resolution}, while in this section, we develop and demonstrate how to enforce this procedure when conflicts emerge between two different \acp{NIF} algorithms acting on the same network functions but configuring different values for the target parameters. During the prototyping, Kubernetes was used as the main deployment infrastructure, while the following phases of the conflict resolution process were addressed: (i) \ac{NIFD} creation and annotation following the \ac{N-MAPE-K} taxonomy; (ii) initial assessment based on \ac{N-MAPE-K} types; (iii) conflict identification and; (iv) conflict resolution. The process is illustrated in Figure~\ref{fig:conflict_resolution_implementation} and explained in the current section.

To demonstrate the conflict resolution capabilities of the \ac{NISt}, two \acp{NIF} were created and deployed. The first \ac{NIF} includes a federated learning-powered anomaly detection algorithm (identical to the \ac{NIF}1 presented in Section~\ref{sec:results:orchestrate_implementation}) extended with a basic service resource management capability (e.g., scale in/out on resource under/ overutilization). The second \ac{NIF} remains identical to the \ac{NIF}2, as explained in the previous section. The configuration actions of the above \acp{NIF} (scale in/out and service relocation) cause conflicts because they may act upon the same service, driving the network to unstable conditions if executed in a similar time window. Therefore, in the prototype, we showcase the identification of the conflict while appropriate policy rules are generated before deployment to avoid unstable conditions by orchestrating the configuration actions.

\begin{figure}
    \centering
    \includegraphics[width=\columnwidth]{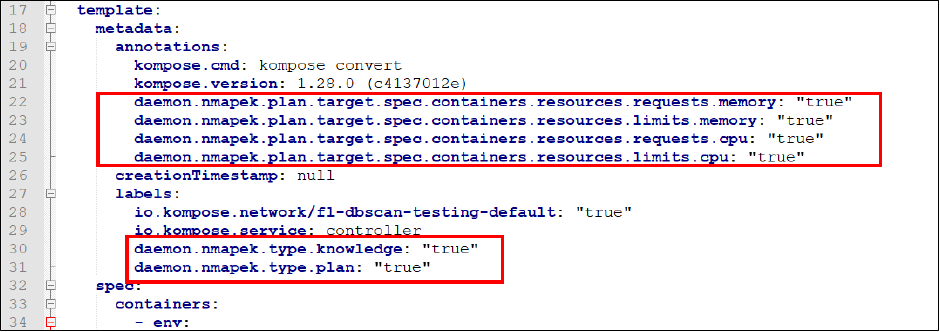}
    \caption{Descriptor of a NIF (NIFD) that can perform service scale out.}
    \label{fig:NIFD1-implementation}
\end{figure}

\begin{figure}
    \centering
    \includegraphics[width=\columnwidth]{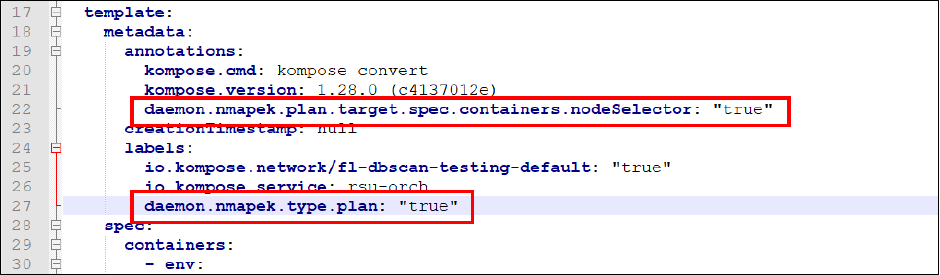}
    \caption{Descriptor of a NIF (NIFD) that can perform service relocation.}
    \label{fig:NIFD2-implementation}
\end{figure}

Initially, the \acp{NIFD} of both \acp{NIF} are created. Since each \ac{NIF-C} is deployed as Pod in Kubernetes, we create the \acp{NIFD} directly from the Kubernetes manifest files. In this direction, the \acp{NIFD} are generated by extending such files in two ways: (i) labels are used to discriminate between different \ac{N-MAPE-K} types (Sensor, Monitor, Analyse, Plan, Execute, Effector); (ii) annotations are used to specify the configuration capabilities of the \ac{NIF} in a self-descriptive way. This process is illustrated in steps (1) and (2) in Figure~\ref{fig:conflict_resolution_implementation}. 

The \acp{NIFD} of the two aforementioned \acp{NIF} are illustrated in Figure~\ref{fig:NIFD1-implementation} and Figure~\ref{fig:NIFD2-implementation} respectively. The first \ac{NIFD} (Figure~\ref{fig:NIFD1-implementation}) is of type ``Knowledge" due to the federated learning-powered anomaly detection capabilities and ``Plan" since it can perform service scale out. In detail, the possible configuration targets follow the prefix ``daemon.nmapek.plan.target" followed by the resource they can configure. This structure is already used in the Kubernetes manifest files (e.g., ``spec. containers.resources.requests.memory"). In the same manner, Figure~\ref{fig:NIFD2-implementation} illustrates the \ac{NIFD} of the second \ac{NIF}. This \ac{NIF} is of type ``Plan" since it can perform service relocation, affecting the ``spec.containers.nodeSelector" part of a Kubernetes manifest file. 

Then, the \ac{CSOI} module is responsible for parsing the provided \acp{NIFD} to identify the \ac{N-MAPE-K} types included in the \ac{NIF} and the possible configuration actions that the \acp{NIF} can generate (step 3 in Figure~\ref{fig:conflict_resolution_implementation}). In addition, an initial assessment is performed to quickly identify if a conflict process is required (step 4). The \ac{CSOI} communicates with the \ac{PolicyIC} module, which forwards only the relevant parameters to the Conflict Resolution module. The Conflict Resolution module executes a thorough analysis to identify any possible conflict (step 6). The analysis is based on a set of conflict resolution matrices and the list of already deployed \acp{NIS}. 

In this direction, a conflict may be identified between an existing \ac{NIS} and a new \ac{NIS} to be deployed. In case of conflicts, the conflict solver selector is triggered. This module selects the appropriate Conflict Solver from a set of available Conflict Solvers (step 7). The Conflict Solver is a software component that can resolve a conflict during the deployment and operation phases. This means that it may update the existing capabilities of a \ac{NIS} or generate Policies as a set of rules (step 8) to handle conflicting situations during the operation of conflicting \acp{NIF}. An example of a rule is also illustrated in Figure~\ref{fig:conflict_resolution_implementation}, where \ac{NIF}2 will wait 30 seconds before relocating the service if \ac{NIF}1 already scaled it. In the case of new Policies, the Policy Manager is responsible for adding them to the Policies database and becoming responsible for keeping them updated during the whole lifecycle. After all checks are made and any Policies are generated and activated, the new \ac{NIS} is deployed by the K8S Deployer (steps 10 and 11).

During operation, all the certified \ac{NIF} coexist conflict-free in the same Kubernetes Cluster (step 11). In the case of a conflicting \ac{NIF}, the candidate conflicting \ac{NIF} communicates with the PolicyIC (and \ac{NIS} Planning Orchestrator) before any conflicting action is realized (step 12). Then, the \ac{NIS} Planning Orchestrator is responsible for retrieving the related Policies from the Policy Manager and applying their rules. After the related Policies are executed, the PolicyIC may or may not give the green light to the \ac{NIF} to complete the requested configuration (step 14).

\section{Conclusion}\label{sec:conclusion}

This paper presents the architectural design of the \ac{NISt}, an evolution of a \ac{NI} plane, an end-to-end orchestrator for \ac{NI}. This design serves multiple purposes: it supports closed-loop \ac{NI} across the entire network infrastructure, enables coordination of \ac{NI} instances to exploit synergies and manage conflicts, and defines necessary interfaces for \ac{NI} algorithms to interact with local environments. This architectural approach is pivotal in structuring \ac{NI} more effectively and is instrumental in enhancing the efficiency and effectiveness of \ac{NI} deployment in network infrastructures.

We build upon our previous work on defining \ac{NI}, highlighting the importance of data and the mechanisms required for coordinating \ac{NI} instances. We have extended this prior work by introducing a comprehensive workflow for managing the \ac{NI} lifecycle. This includes detailed procedures for resolving conflicts among multiple \ac{NI} instances and strategies for knowledge sharing. The methodology outlined in the paper is an important step forward in the field, addressing key challenges in \ac{NI} coordination and conflict resolution.

The research presented in the manuscript not only extends theoretical and methodological aspects of \ac{NI} but also focuses on practical applicability. The paper expands the scope of the reference implementation to include necessary modifications that support conflict resolution capabilities in leading open-source platforms such as Kubernetes, Kubeflow, and Zenoh. This expansion is crucial for broadening the potential impact and utility of the research, making it more relevant and applicable in real-world scenarios.

\subsection{Open Challenges}
Deploying and managing a \ac{NISt} is a complex process that involves the integration and coordination of multiple frameworks, not only including network \ac{MANO} and \ac{MLOps} frameworks, as shown in this paper, but also data management frameworks~\cite{zeydan2022recent}. Moreover, developing production-ready \ac{NI} models is cumbersome, comprising the optimization of complex networking tasks and hyperparameter tuning, all while adhering to network constraints such as throughput and latency. This complexity hinders the efficient and error-free deployment and management of \ac{NI} and calls for proficiency not only in \ac{ML} but also in network design and programmable hardware.  

An ongoing challenge in the deployment of the \ac{NISt} involves the detection and resolution of conflicts, especially when multiple \ac{NI} operate within different network domains. The need for effective policy interpretation, configuration, and enforcement to manage these conflicts is crucial. This requires a robust mechanism for interpreting \ac{NI} decisions and outcomes. Besides improving trustworthiness in black-box \ac{NI} models, \ac{XAI}~\cite{brik2023survey} can offer insights into the inner workings of \ac{NI} models, thereby demystifying their outcomes. Moreover, as networks become increasingly autonomous, explaining decisions by the \ac{NI} becomes imperative, especially in scenarios where the decisions may have critical consequences. \ac{XAI} provides a framework for ensuring that \ac{AI}-driven decisions within \ac{NI} are fair, unbiased, and compliant with regulatory standards.

Although out of scope in this paper, developing accurate network digital twins is also a paramount direction for future research in NI. These abstractions help to bridge the gap between \ac{ML} model training and deployment in autonomous networks. During training, the \ac{NI} model parameters are fine-tuned to suit the particular challenges the model is intended to solve. To mitigate the risk of introducing instability to the network due to premature decisions, the outputs from the training phase should be first tested within a digital twin of the network. Such a digital twin acts as a safe environment, allowing potential errors or inefficiencies from an incomplete training process to be identified and rectified without any adverse impact on the actual network operations, as suggested by Almasan et al.~\cite{almasan2022network}. The challenge lies in creating a digital twin that precisely mirrors the complexities and dynamics of the target network, which remains an open issue in the field.

\section*{Acknowledgment}
This work has received funding from the European Union’s Horizon 2020 research and innovation program under grant agreement No.~101017109 ``DAEMON''.

\bibliographystyle{elsarticle-num}

\end{document}